\documentclass[onecolumn,pr]{revtex4}
\usepackage{verbatim}
\usepackage{amsmath,amssymb}
\usepackage{graphicx}
\usepackage{xcolor}
\usepackage[colorlinks,bookmarks=false,citecolor=blue,
linkcolor=red,urlcolor=blue]{hyperref}
\usepackage{times}
\usepackage{soul}

\def \beq{\begin{equation}}
\def \eeq{\end{equation}}
\def \barray{\begin{eqnarray}}
\def \earray{\end{eqnarray}}

\font\numbers=cmss12
\font\upright=cmu10 scaled\magstep1
\def\stroke{\vrule height8pt width0.4pt depth-0.1pt}
\def\topfleck{\vrule height8pt width0.5pt depth-5.9pt}
\def\botfleck{\vrule height2pt width0.5pt depth0.1pt}
\def\Zmath{\vcenter{\hbox{\numbers\rlap{\rlap{Z}\kern
0.8pt\topfleck}\kern 2.2pt
                   \rlap Z\kern 6pt\botfleck\kern 1pt}}}
\def\Qmath{\vcenter{\hbox{\upright\rlap{\rlap{Q}\kern
                   3.8pt\stroke}\phantom{Q}}}}
\def\Nmath{\vcenter{\hbox{\upright\rlap{I}\kern 1.7pt N}}}
\def\Cmath{\vcenter{\hbox{\upright\rlap{\rlap{C}\kern
                   3.8pt\stroke}\phantom{C}}}}
\def\Rmath{\vcenter{\hbox{\upright\rlap{I}\kern 1.7pt R}}}

\begin{document}

\author{Vincenzo Alba$^1$, Silvia N.~Santalla$^2$, Paola Ruggiero$^1$, Javier Rodriguez-Laguna$^3$, Pasquale Calabrese$^{1,4}$, and German Sierra$^5$}

\affiliation{$^1$International School for Advanced Studies (SISSA) and INFN,
Via Bonomea 265, 34136, Trieste, Italy}
\affiliation{$^2$Departamento de F\'isica and Grupo Interdisciplinar de Sistemas Complejos (GISC), Universidad Carlos III de Madrid, Spain}
\affiliation{$^3$Departamento de F\'isica Fundamental, Universidad Nacional de Educacion a Distancia (UNED), Madrid, Spain}
\affiliation{$^4$International Centre for Theoretical Physics (ICTP), 34151, Trieste, Italy}
\affiliation{$^5$Instituto de F\'isica Te\'orica UAM/CSIC, Universidad Aut\'onoma de Madrid, Madrid, Spain}

\date{July 4, 2018}

\title{Unusual area-law violation in random inhomogeneous systems}  

\begin{abstract} 
The discovery of novel entanglement patterns in quantum many-body systems 
is a prominent research direction in contemporary physics. 
Here we provide the example of a spin chain with random and inhomogeneous couplings
that in the ground state exhibits a very unusual area law violation. 
In the clean limit, i.e., without disorder, the model is the rainbow chain and has volume law entanglement. 
We show that, in the presence of disorder, the entanglement entropy exhibits a 
power-law growth with the subsystem size, with an exponent $1/2$. 
By employing the Strong Disorder Renormalization Group (SDRG) framework, we show 
that this exponent is related to the survival probability of certain random walks. 
The ground state of the model exhibits extended regions of short-range singlets (that we term ``bubble'' regions) as 
well as rare long range singlet (``rainbow'' regions). 
Crucially, while the probability of extended rainbow regions decays exponentially with their size, that of the 
bubble regions is power law. 
We provide strong numerical evidence for the correctness of SDRG results by exploiting the free-fermion solution of the model. 
Finally, we investigate the role of interactions by considering the random inhomogeneous XXZ spin chain. 
Within the SDRG framework and in the strong inhomogeneous limit, we show that the above area-law violation 
takes place only at the free-fermion point of phase diagram. 
This point divides two extended regions, which exhibit volume-law and area-law entanglement, respectively. 
\end{abstract}


\maketitle

\section{Introduction}
\label{intro}

\begin{figure}[t]
\includegraphics*[width=0.75\linewidth]{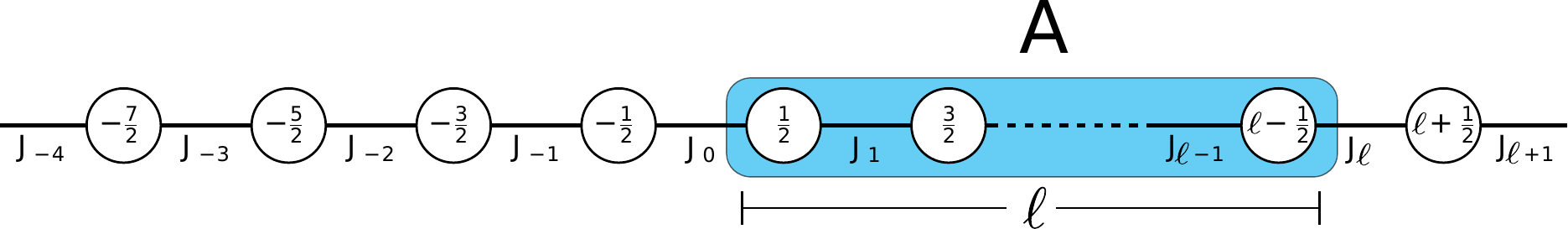}
\caption{Setup used in this work. (top) Definition of the 
 random inhomogeneous XX chain. The chain couplings are 
 denoted as $J_{n}=e^{-|n|h}K_n$, with $n$ half integer numbers, 
 $h$ a real inhomogeneity parameter, and $K_n$ independent random 
 variables distributed with~\eqref{K-distr}. In this work we 
 focus on the entanglement entropy of a subregion $A$ of 
 length $\ell$ (shaded area in the Figure). Subsystem 
 $A$ starts from the chain center. 
}
\label{fig0}
\end{figure}

A striking feature of local gapped quantum many-body systems is that the 
ground-state entanglement entropy of a subsystem scales with the area of its 
boundary rather than with its volume~\cite{area,amico-2008,calabrese-2009,laflorencie-2016}. 
This statement is the essence of the famous {\it area law} for the entanglement. 
Given a quantum system in a pure state in $D$ dimensions, and given a bipartition 
of the system into a subsystem $A$ and its complement $\bar A$ (see, for instance, 
Figure~\ref{fig0} for a one-dimensional setup), the von Neumann entanglement entropy is defined 
as
\begin{equation}
S\equiv-\textrm{Tr}\rho_A\ln\rho_A, 
\end{equation}
where $\rho_A$ is the reduced density matrix of $A$, which is obtained 
by tracing over the degrees of freedom of $\bar A$ in the full-system 
density matrix $\rho$. After denoting as $\ell$ the typical length 
of $A$, the area law states that for large $\ell$ the entanglement entropy 
scales like $S\propto\ell^{D-1}$. 
Physically, the area law suggests that the ground state of local hamiltonians 
contains much less quantum correlation than what one might have expected. 
The area law has enormous consequences for the simulability of quantum states 
using classical computers. For instance, it  underlies the extraordinary 
success of Matrix Product States (MPS) methods, such as the Density Matrix 
Renormalization Group~\cite{white1,white2,uli} (DMRG) to effectively describe ground states of 
one-dimensional systems. 
For gapped many-body systems, there is an unanimous consensus that the area 
law is valid  in arbitrary dimension~\cite{area}, although a rigorous 
proof is only available for one-dimensional systems~\cite{hastings-2007} (see 
also \cite{arad-2013}). 
Conversely, it is well known that the ground states of gapless free-fermionic hamiltonians exhibit 
logarithmic corrections to the area law~\cite{wolf-2006,gioev-2006,cmv-11d}, i.e., one has $S={\mathcal O}
(\ell^{D-1}\ln\ell)$, in contrast with gapless bosonic systems~\cite{plenio-2005}, 
for which no corrections are present for $d\geq2$. 
However, the most prominent examples of logarithmic area-law violations 
are critical one-dimensional models whose low energy properties are captured by a 
Conformal Field Theory (CFT)~\cite{holzhey-1994,calabrese-2004,vidal-2003,cc-09}, 
and spin chains with a permutation symmetric ground state~\cite{popkov-2005,ercolessi-2011,
alba-2012,olalla-2012}. Importantly, the area-law is not generic. 
{\it Typical} excited states of local hamiltonians exhibit a volume law 
entanglement~\cite{gltz-05,hayden-2006,alba-ex} (and in these cases the density 
of entanglement entropy is the same as the thermodynamic entropy of a 
generalised microcanonical ensemble at the correct energy, see, e.g.,  \cite{alba-2016}).
However, there are many examples of eigenstates with sub-volume (logarithmic) scaling of the entanglement entropy 
(see, e.g., Refs.~\cite{alba-ex,cdds-18}), in particular when the low-energy part of the spectrum is described by a 
CFT for which exact analytic predictions are obtainable \cite{abs-11,sierra2012,p-14,p-16,txas-13,top-16,elc-13,CEL,cmv-11b}. 

Motivated by this evidence, there is strong common belief that ground states of 
``physically reasonable local hamiltonians'' fulfil the area law, and that 
violations are at most logarithmic (see, however, Ref.~\cite{farkas-2005}). 
Only very recently, devising local models that exhibit more dramatic area-law 
violations became an important research theme. 
The motivation is twofold. On the one hand, highly-entangled 
ground states are potentially useful for quantum computation technologies. 
On the other hand, from a condensed matter perspective, area-law 
violations could be witnesses of exotic features of quantum matter. 
As a matter of fact, examples of ground states violating the area law 
start to be discovered (see, for instance,~\cite{irani-2010,gottesman-2010,vitagliano-2010,
bravy-2012,lisa-2013,shiba-2014,gori-2015,mova-2016,salberger-2016a,zhang-2017,salberger-2017,sugino-2017,caha-2018,korepin-l}). 
These comprise inhomogeneous systems~\cite{vitagliano-2010}, translation 
invariant models with large spin~\cite{irani-2010}, 
free-fermion hamiltonians~\cite{gori-2015} with a fractal Fermi surface, 
nonlocal Quantum Field Theories~\cite{shiba-2014}, 
and supersymmetric models~\cite{lisa-2013}. 
An interesting class of frustration-free, local, translational invariant models that exhibit area-law violations 
has been constructed by Movassagh and Shor in Ref.~\cite{mova-2016}. 
Their  ground-state entanglement entropy is $\propto\ell^{1/2}$, thus exhibiting 
a polynomial violation of the area law. Importantly, the exponent of 
the entanglement growth originates from universal properties of the random 
walk. This is due to the fact that the ground state of the model is written 
in terms of a special class of combinatorial objects, called Motzkin 
paths~\cite{motzkin}. A similar result can be obtained~\cite{salberger-2016a} 
using the Fredkin gates~\cite{fredkin}. An interesting generalization of 
Ref.~\onlinecite{mova-2016} obtained by deforming a colored version of the 
Motzkin paths has been presented in Ref.~\onlinecite{zhang-2017}. 
The ground-state phase diagram of the model exhibits two phases 
with area-law and volume-law entanglement, respectively. These 
are separated by a ``special'' point, where the ground state 
displays square-root entanglement scaling. 

%
%

In this paper, we show that unusual area-law violations can be obtained 
in a one-dimensional inhomogeneous local system in the presence of disorder. 
Specifically, here we investigate the random inhomogeneous XX chain. 
In the clean limit, i.e., in the absence of disorder, our model reduces to the rainbow chain of Ref.~\cite{vitagliano-2010}, 
whose ground state, in the limit of strong inhomogeneity, is the rainbow state.
In the rainbow state long-range singlets are formed between spins across the chain center. 
An immediate striking consequence is that the half-chain entanglement entropy  is proportional to the 
subsystem volume (volume law). 
Here we show that upon including disorder the structure of the ground state changes dramatically. 
In contrast with the clean case, now the probability of having 
long-range singlets across the chain center is strongly suppressed. 
In particular, the probability of having extended regions (that we 
term ``rainbow'' regions) of mirror symmetric singlets across the chain center decays 
exponentially with the region size. 
On the other hand, the probability of having extended regions with short-range singlets 
connecting nearest-neighbor spins decays algebraically with 
the region size, with an exponent $3/2$. 
This has striking consequences for the entanglement scaling. Precisely, in contrast with the clean 
case, the entropy exhibits an unusual square root growth, which 
represents a polynomial violation of the area law. 
We provide numerical evidence for this behavior by using the Strong Disorder 
Renormalization Group (SDRG) method~\cite{igloi-rev} (see also\cite{igloi-rev-1}). 
We numerically verify that the unusual area law violation happens both in the 
strong inhomogeneous limit, as well as for weak inhomogeneity. 
Specifically, we numerically observe 
the square-root scaling for considerably weak inhomogeneity, 
although we do not have any proof that it persists for arbitrarily small 
one. We provide robust numerical evidence of 
the unusual area-law violation in a microscopic model by calculating 
the entanglement entropy of the random inhomogeneous XX chain, which 
is obtained by using the free-fermion solution of the model. Furthermore, 
 we establish a mapping 
between the SDRG flow of the renormalized couplings and an alternating 
random walk. Interestingly, in the strong inhomogeneous limit the exponents of the entanglement scaling, and 
several ground-state features, can be quantitatively understood 
from certain survival probabilities of the random walk. Finally, we 
investigate the role of interactions by considering the random inhomogeneous 
spin-$1/2$ XXZ chain. Within the SDRG framework, we show that the unusual 
area-law violation does not survive in the presence of interactions. 
Precisely, we find 
that the entanglement entropy exhibits square-root scaling only at the 
XX point. Interestingly, this marks the transition between two extended 
regions, where the entanglement entropy exhibits area-law and volume-law 
scaling, respectively. Both the two behaviors can be qualitatively 
understood in the SDRG framework by exploiting the mapping to the random walk, at least in the 
strong inhomogeneous limit.  This
scenario is somewhat similar to the one presented in Ref.~\onlinecite{zhang-2017}, 
although the models are substantially different. 

The paper is organized as follows. In section~\ref{sec-model} 
we introduce the random inhomogeneous XX chain and the SDRG framework. 
In particular, in section~\ref{sec-lemma} we show that for the XX chain 
the SDRG renormalization flow exhibits an intriguing independence on 
the specific renormalization pattern. In section~\ref{sec-area} we 
present numerical SDRG results for the entanglement entropy in the 
XX chain. In section~\ref{sec-cont} we describe how the unusual 
area-law violation is reflected in the behavior of the  
entanglement contour. In section~\ref{sec-XX}, by exploiting the 
exact solvability of the random inhomogeneous XX chain, we provide 
evidence of the area-law violation. In sections~\ref{sec-toy} 
 we address the strongly inhomogeneous limit 
of the model, by exploiting a mapping between the SDRG flow and 
the random walk. Section~\ref{sec-int} is devoted to discuss the 
entanglement scaling in the XXZ chain. We conclude in 
section~\ref{sec:conc}. Finally, 
in Appendix \ref{sec-algebra} we propose  an algebraic
interpretation of the SDRG scheme, and 
in Appendix~\ref{sec-calc} we 
report the calculations of certain survival probabilities 
for the random walk introduced in~\ref{sec-toy}.

\section{The random inhomogeneous XX chain (randbow chain)} 
\label{sec-model}

We consider a chain with $2L$ sites, described by the following 
inhomogeneous random hopping hamiltonian (see Figure~\ref{fig0})
%
%
%
\begin{equation}
\label{ham}
H= -  \frac{1}{2} \sum\limits_{m=- L+1}^{L-1}
J_m  \; c^\dagger_{m- \frac{1}{2} } c^{}_{m+ \frac{1}{2}}  +\textrm{h.c.},\quad
\textrm{with}\,\,m= 0, \pm 1, \pm 2, \cdots , \pm (L- 1) .
\end{equation}

Here $c_{m \pm \frac{1}{2}}$ ($c^{\dagger}_{m \pm \frac{1}{2}}$) denotes the annihilation (creation) operator 
of a spinless fermion at sites $m \pm \frac{1}{2}$, and $J_m>0$ is the  inhomogeneous random 
hopping parameter between the sites $m- \frac{1}{2}$ and $m + \frac{1}{2}$. 
In~\eqref{ham}, the coupling $J_0$ is associated to the link $\left( - \frac{1}{2}, \frac{1}{2} \right)$ located at 
 the center of the chain. 
The hopping parameters $J_m$ are 
defined as 
%
%
%
\begin{equation}
\label{coup}
J_m\equiv K_m  \times \left\{ 
\begin{array}{ll} 
e^{ - h/2} , & m=0 \, ,  \\
e^{ - h  |m| }, & |m| >  0 \, ,  \\
\end{array}
\right. 
\end{equation}

where $h>0$ is a real parameter that measures the strength of the 
inhomogeneity. If $K_m={\mathcal O}(1)$ are nonzero, 
for $h>0$ the coupling strength decreases exponentially with the distance 
from the chain center. In~\eqref{coup}, we choose $K_m$ to be independent 
(from site to site) random variables distributed  in the interval $[0,1]$ 
according to
\begin{equation} 
\label{K-distr}
P(K) = \delta^{-1} K^{-1 + \frac{1}{\delta}},
\end{equation}
with $\delta>0$ parametrizing the noise strength. For $\delta=1$, $P(K)$ becomes 
the uniform distribution in the interval $[0,1]$. For $\delta\to 0$, $P(K)$ is 
peaked at $K=1$ and the model~\eqref{ham} is clean, i.e., without disorder. On 
the other hand, for $\delta\to\infty$, $P(K)$ is peaked at $K=0$. In the limit 
$\delta\to\infty$, Eq.~\eqref{K-distr} defines the  Infinite Randomness 
Fixed Point (IRFP) distribution, which describes the long-distance properties 
of the ground state of~\eqref{ham} for $h=0$ and any $\delta$ (see below). 

After a Jordan-Wigner transformation, the random hopping model in~\eqref{ham} is 
mapped onto the spin-$1/2$ inhomogeneous XX chain defined by 
%
%
%
\begin{equation}
\label{ham-XX}
H= \frac{1}{2}\sum\limits_{m= -L+ 1}^{L-1}
J_m \; S_{m - \frac{1}{2}} ^+S^-_{m+ \frac{1}{2}}   +\textrm{h.c.},\quad
\textrm{with} \;  \,m= 0, \pm 1, \pm 2, \cdots , \pm (L- 1)  \, . 
\end{equation}

Here $S_m^\pm$ are spin-$1/2$ raising and lowering operators.

%
\begin{figure}[t]
\includegraphics*[width=\linewidth]{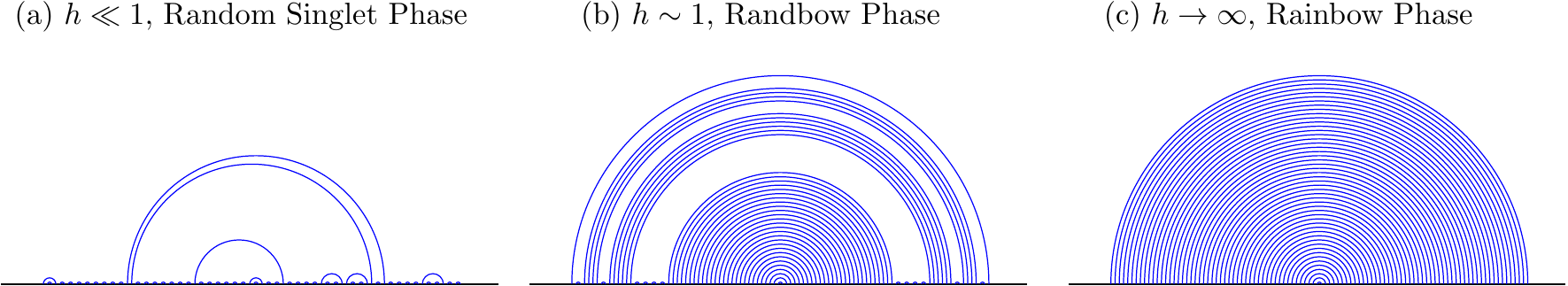}
\caption{ Summary of the phase diagram of the model, using open boundaries. Arcs in the Figure correspond to spins forming $SU(2)$ singlet bonds. Notice that for any $h\ne 0$ the coupling strength $J_i$ decreases exponentially away from the center of the chain.
(a) For $h\ll 1$ the model becomes the XX chain with random antiferromagnetic couplings. The ground state of the model is the random singlet phase (RSP). In the RSP bonds of arbitrary length are present, but no symmetry with respect to the chain center is observed. RSP phases exhibit logarithmic entanglement
growth.
(b) For intermediate values of $h$ we observe some long distance bonds along with a proliferation of short ones, connecting neighboring sites (bubbles). The bond diagram presents left-right symmetry, and the entanglement is characterized by a subextensive (square root) entanglement growth.
(c) For $h\to\infty$ the model approaches the standard rainbow chain, with all bonds symmetric with respect to the chain center and exhibiting volume-law entanglement.} 
\label{fig0-a}
\end{figure}
%

In this work we investigate the ground-state entanglement entropy $S$ of a subregion $A$ 
that starts from the chain center. The precise bipartition that we consider 
is pictorially illustrated in Figure~\ref{fig0}. 

Clearly, the properties of the model~\eqref{ham-XX} depend on two parameters, 
$h$ and $\delta$, giving rise to a two-dimensional ground-state phase diagram. 
The clean homogeneous XX chain is recovered for $\delta\to 0$ and $h=0$. Its 
ground state is critical, and it is described by a Conformal Field Theory (CFT) 
with central charge $c=1$. The entanglement entropy of a finite subsystem $A$ 
exhibits a logarithmic area-law violation described by  \cite{cc-09}
\begin{equation}
\label{cft}
S=\frac{c}{3}\ln\ell+k, 
\end{equation}
with $k$ a non-universal constant, $\ell$ the size of $A$, and $c=1$ the central 
charge of the CFT. 
In this work we focus on the entanglement properties of~\eqref{ham-XX} 
at $\delta>0$. In the limit $h\to0$ and for any finite $\delta$ (cf.~\eqref{coup}), 
Eq.~\eqref{ham-XX} defines the random antiferromagnetic XX chain. The ground state 
of the model has been extensively studied using the Strong Disorder Renormalization 
Group~\cite{ma-1980,igloi-rev,fisher} (SDRG), and it is described by the random singlet (RS) phase. The structure of 
the random singlet phase is depicted in Figure~\ref{fig0-a} (a). In the Figure the 
links denote a singlet bond between the spins at their end points. In the RS phase 
short bonds between spins on near neighbour sites are present, as well as bonds 
joining distant spins. A distinctive feature of the RS phase, which can be 
derived by using the SDRG approach, is that, similar 
to the clean case (cf.~\eqref{cft}), the entanglement entropy of a finite 
subregion scales logarithmically as~\cite{refael-2004,refael-2009,laflorencie-2005,fagotti-2011} 
\begin{equation}
\label{rsp-ent}
S=\frac{\ln 2}{3}\ln\ell+k',
\end{equation}
where $k'$ a non universal constant. The crucial difference with~\eqref{cft} 
is the $\ln 2$ prefactor. For the random XX chain, this prefactor can be 
interpreted as a renormalization of the central charge $c=1$ due to the 
disorder, which reduces the amount of the entanglement (see, however, 
Ref.~\onlinecite{raoul-2006} for a counterexample). 

For $\delta=0$ and $h>0$, the ground state of~\eqref{ham-XX} is in the 
rainbow phase~\cite{vitagliano-2010}. The structure of the ground state of~\eqref{ham-XX} is 
illustrated in Figure~\ref{fig0-a} (c). The system exhibits a proliferation of 
long bonds connecting distant spins symmetrically across the chain center. 
This behavior can be understood in the strong inhomogeneity limit at $h\gg1$, 
By using the SDRG method, one can easily show that the ground state 
of~\eqref{ham-XX} is the \emph{rainbow} state. In the language of fermions this 
reads 
\begin{equation} 
\label{rainbow-state}
|\text{RAINBOW}\rangle = \prod_{n= 1/2}^{L- 1/2} \left( 
c^{\dagger}_{-n} + (-1)^{n-1/2}  c^{\dagger}_{+n} \right) |0 \rangle.
\end{equation}
In the spin representation, the state~\eqref{rainbow-state} corresponds to 
a product of singlets between the sites $(-n, +n)$ of the chain. An 
important feature of the rainbow state~\eqref{rainbow-state} is that 
the entanglement entropy of a subsystem starting from the chain center grows 
linearly with its size $\ell$ (corresponding to a \emph{volume} law) 
as~\cite{vitagliano-2010,ramirez-2014b,ramirez-2015,ramirez-2014a,laguna-2016,laguna-2017}
\begin{equation} 
\label{vol}
\lim_{h \to \infty} S(h, \ell) = \ell\ln 2, 
\end{equation}
where subleading ${\mathcal O}(1)$ terms have been neglected. Eq.~\eqref{vol} 
reflects that the entanglement is proportional to the number of singlets 
shared between $A$ and its complement $\bar A$, i.e., connecting a site in $A$ 
and the other in $\bar A$. Remarkably, the volume-law scaling~\eqref{vol} 
survives in the weak inhomogeneity limit $h \to 0$. One can take the continuum 
limit of~\eqref{ham}, by sending the lattice spacing $a\to0$ and by considering  
$h\to0$ and $L\to\infty$ with $h/a$ and $aL$ fixed, to show that the half-chain 
entanglement entropy is still linear with $L$, but with a different coefficient 
as~\cite{laguna-2017} 
\begin{equation}
\label{c-cft}
S(h, L)\simeq \frac{1}{6} \ln\left( \frac{e^{hL} -1}{h} \right) \to \frac{h 
L }{6}.
\end{equation}
The last expression in~\eqref{c-cft} is obtained in the limit $hL \gg 1$.

In this work we focus on the regime with finite nonzero $\delta$ and $0<h<\infty$. 
In this regime the ground state of~\eqref{ham-XX} is in a dramatically different phase. 
This is illustrated in Figure~\ref{fig0-a} (b). 
Its structure is easily understood in the limit $h\gg1$. 
Similar to the rainbow phase, long bonds connecting spins on symmetric sites 
with respect to the center of the chain are present. However, in contrast with the 
rainbow case (see Figure~\ref{fig0-a} (c)), they are rare and do not 
form an extended phase. Precisely, 
the probability of forming a sequence of rainbow links decreases exponentially with 
its size, i.e., with the number of consecutive sites involved. On the other hand, 
the ground state of~\eqref{ham-XX} exhibits a proliferation of short-range singlets 
between spins on nearest-neighbor sites. These form extended ``bubble'' regions 
(see Figure~\ref{fig0-a} (b)). We anticipate that the probability of forming a 
bubble region of length $\ell_b$ decays as a power law as $\propto 
\ell_b^{\scriptscriptstyle-3/2}$, 
in contrast with that of forming a rainbow region, which is exponential. 
This has striking consequences for the scaling of the entanglement entropy. 
First, only rainbow bonds can contribute to the entanglement between $A$ and 
the rest, because short singlets connect mostly sites within $A$ and $\bar A$, 
separately.  On the other hand, the typical length scale over which the system 
is entangled is determined by the scaling of the regions with short-range singlets. 
Specifically, our main result is that for $0<h<\infty$ and finite $\delta$ 
the von Neumann entropy exhibits a square-root scaling behavior as 
\begin{equation}
\label{sq-ent}
S=C\cdot\ell^{1/2}+k'',
\end{equation}
where $C$ and $k''$ are non-universal constants. Notice that 
Eq.~\eqref{sq-ent} represents a dramatic violation of the area-law. 

\subsection{Strong Disorder Renormalization Group (SDRG) method}
\label{sec-sdrg}

Away from the limits $h=\delta=0$, the hamiltonian~\eqref{ham-XX} can be 
studied using the SDRG technique, first introduced by Dasgupta and Ma~\cite{ma-1980}. 
In the standard SDRG framework, high-energy degrees of freedom in~\eqref{ham-XX} 
are progressively removed from the model via a decimation procedure. This works  
as follows. At each SDRG step, we select the link with the largest value of the 
coupling $J_M\equiv\max\{J_i\}$ (cf.~\eqref{coup}). Thus, we put in a singlet state the two spins 
connected by the link. This has the effect of renormalizing the interaction between 
the next-nearest neighbor spins. This effect can be derived by treating the couplings 
on the links next to $J_M$ using standard second-order perturbation theory. The 
resulting effective coupling $J'$ between the next-nearest neighbor spins is 
obtained as 
\begin{equation}  
\label{RG-equation}
J' =\frac{J_L J_R}{J_M},
\end{equation}
where $J_L$ and $J_R$ are the coupling to the left and to the right of $J_M$, respectively. 
After many iterations of the SDRG step, all the spins are decimated, and the 
resulting state is a collection of singlets, i.e., a valence bond state (VBS). 

It is useful to rewrite the SDRG procedure by introducing the logarithmic couplings $T_m$ as 
\begin{equation}
T_m\equiv-\ln J_m. 
\end{equation}
Notice that $T_m$ takes into account both the random part of the coupling $K_m$ 
(cf.~\eqref{coup}), as well as the inhomogeneity due to the presence 
of $h$. Notice that the contribution of $h$ is a non-random position-dependent 
shift in $T_m$.
In the variables $T_m$ the SDRG renormalization step~\eqref{RG-equation} 
becomes additive. We anticipate that this allows one to interpret the SDRG 
procedure as a random walk in the space of $T_m$ (see, 
for instance, section~\ref{sec-lemma}).

The structure of the SDRG renormalization has been intensively investigated for the 
homogeneous random~\cite{igloi-rev} XX chain which is obtained for $h=0$. As it was anticipated, 
the ground state of the model is described by the  random singlet 
(RS) phase. In the RS phase singlets are mostly formed between nearest-neighbor 
sites, although random singlets connecting spins at arbitrary large distance are also 
present. Although they are suppressed, the latter are responsible for a slow decay, 
as a power law, of the spin-spin correlation function.  
%
%
An important observation is that after many SDRG steps the distribution of the 
renormalized couplings is of the form~\eqref{K-distr} with $\delta\to\infty$. 
This implies that the strength of the disorder effectively increases during the SDRG flow, 
which justifies the use of perturbation approach in~\eqref{RG-equation}, and the 
applicability of the SDRG method. Finally, we should mention that the SDRG 
approach proved to be the method of choice to understand the entanglement 
scaling in generic disordered systems~\cite{igloi-rev,lin-2007,igloi-2012,hoyos-2007,juh-2007,
igloi-2009,hoyos-2011,kovacs-2012,vosk-2013,juh-2014,getlina-2016,ruggiero-2016,juh-2017,juh-last}. 


Here we choose $K_m$ (see~\eqref{coup}) distributed according 
to~\eqref{K-distr}. Writing this quantity as $K_m = \xi_m^\delta$, one can easily verify that
$\xi_m$ is a random variable uniformly distributed in the interval $[0,1]$. 
This allows one to rewrite the couplings $T_m$ as  
\begin{equation}
\label{coup2}
T_m=\left\{\begin{array}{cc}
h\big(\frac{1}{2}-\frac{\delta}{h}\ln\xi_0\big),  & m=0 \, , \\\\
h\big(|m|-\frac{\delta}{h}\ln\xi_m\big),    & |m|>0 \, . 
\end{array}\right.
\end{equation}
%
An important consequence 
of~\eqref{coup2} is that apart from the overall factor  $h$, the couplings $T_m$ 
are functions of the ratio $\delta/h$ only. Hence,  the VBS state obtained at the end
of the SDRG, as well as the entanglement entropy, only depends on $\delta/h$.


\subsection{Path invariance of the SDRG for the XX chain: A useful lemma}
\label{sec-lemma}

\begin{figure}[t]
\includegraphics*[width=\linewidth]{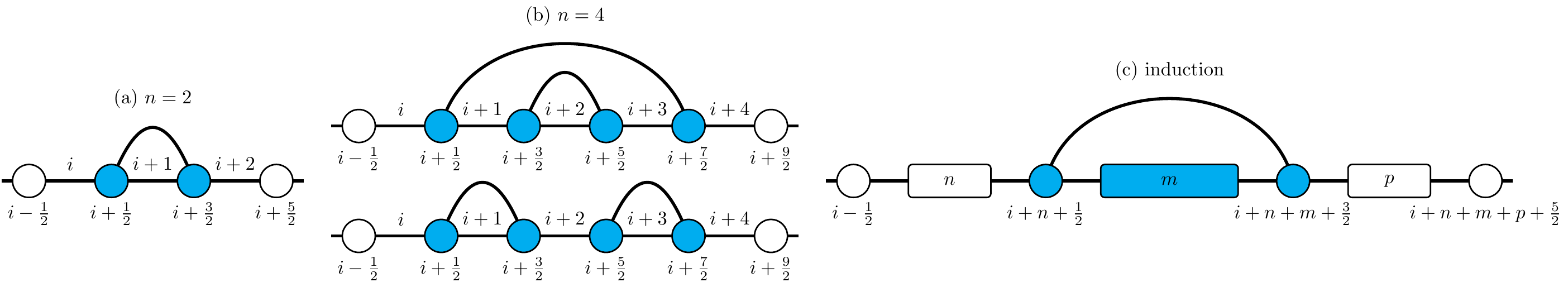}
\caption{Path invariance theorem for the SDRG in the random 
 XX chain: Pictorial proof by induction. The Figure shows that the 
 renormalized coupling between two sites $i- \frac{1}{2}$ and $i+n + \frac{1}{2}$ obtained 
 after decimating all the $n$ spins in between does not depend on the 
 decimation pattern. (a) The theorem for  $n=2$. Only one 
 decimation pattern is possible. (b) $n=4$. Now two patterns 
 are possible: A rainbow and a bubble pattern (top and bottom, 
 respectively). (c) Induction step to prove the theorem. The boxes 
 denote the renormalized couplings for which the theorem applies. 
}
\label{fig-theo}
\end{figure}

Since in this paper we mostly focus on the random XX chain, here we wish to 
discuss a crucial simplification that occurs when one applies the SDRG 
method to this model. We show that for the random XX 
chain (cf.~\eqref{ham-XX}) the renormalized coupling between two sites separated by an odd  number of consecutive bonds is independent of the decimation pattern, and it has a simple form that we provide. To the best of our knowledge this interesting property has not been noticed before in the literature. 
In Appendix \ref{sec-algebra} we propose  an algebraic
derivation of this property in terms of a triplet product inspired by the SDRG method.

Let us start with  the SRDG decimation of the block of four spins shown in 
Figure~\ref{fig-theo} (a), that we denote as $[i- \frac{1}{2}, i + \frac{5}{2}]$, where
$i$ is an integer. The renormalization of this block amounts to the formation
of a bond between the sites $i + \frac{1}{2}$ and $i + \frac{3}{2}$ and a new effective
coupling  between the sites $i - \frac{1}{2}$ and $i + \frac{5}{2}$ whose value is given by
Equation ~\eqref{RG-equation}, 
\begin{equation}
J_{[i- \frac{1}{2},i+ \frac{5}{2}]}=  \frac{J_{i} \; J_{i+2}}{J_{i+1}}  \, .
\end{equation}
Working with the logarithmic couplings $T_i = - \ln J_i$, this equation becomes 
\begin{equation}
\label{theo-2}
T_{[i- \frac{1}{2},i+ \frac{5}{2}]}=T_{i}-T_{i+1}+T_{i+2} \, . 
\end{equation}
The next example involves the renormalization of the block of six consecutive sites $[i - \frac{1}{2}, i + \frac{9}{2}]$
depicted in Figure~\ref{fig-theo}(b). Now there are two possible 
patterns. In the first one (Figure~\ref{fig-theo} (b) top) a nested rainbow diagram 
with two bonds is formed. The other possibility is to create two bubble diagrams 
forming the bonds $(i+ \frac{1}{2}, i+ \frac{3}{2})$  and  $(i+ \frac{5}{2}, i+ \frac{7}{2})$
(Figure~\ref{fig-theo}(b) bottom). Using~\eqref{RG-equation}, it is straightforward 
to check that both decimation patterns give the same effective coupling 
between sites $i- \frac{1}{2}$ and $i+ \frac{9}{2}$, which reads
\begin{equation}
\label{theo-3}
T_{[i- \frac{1}{2},i+ \frac{9}{2}]}=
T_i - T_{i+1}+ T_{i+2} -  T_{i+3} +  T_{i+4} \, . 
\end{equation}
The general expression for the renormalized coupling for a block 
$[i - \frac{1}{2}, i + n +  \frac{1}{2}]$ with $n+1$ bonds is given by 
\begin{equation}
\label{theo-res}
T_{[i- \frac{1}{2},i+n +  \frac{1}{2}]}=\sum_{j=0}^{n}(-1)^j\, T_{i+j} , 
\end{equation}
where $n$ is the number of spins decimated, that must be an even number. 
Equations ~\eqref{theo-2} and ~\eqref{theo-3}
correspond to the cases $n=2$ and $n=4$ of ~\eqref{theo-res} respectively.

We now 
prove~\eqref{theo-res} by general induction. The key step of the proof is summarized 
in Figure~\ref{fig-theo} 
(c). The boxes in the Figure denote renormalized couplings. The numbers 
$n,m,p$ inside the boxes denote the numbers of spins that have been decimated. To proceed by 
induction, we assume that Eq.~\eqref{theo-res} holds for these 
renormalized couplings. Now, two spins are left at positions $i+n+ \frac{1}{2}$ 
and $i+n+m+ \frac{3}{2}$. Without loss of generality we can assume that these 
are the two spins that are decimated 
at the next SDRG step. After the decimation, one obtains that 
the renormalized coupling connecting sites $i- \frac{1}{2}$ and $i+n+m+p+ \frac{5}{2}$ is given as

\begin{equation}
\label{ind-fin}
T_{[i- \frac{1}{2},i+n+m+p+ \frac{5}{2}]} =
T_{[i- \frac{1}{2},i+n+ \frac{1}{2}]} -
T_{[i+n+\frac{1}{2},i+n+m+ \frac{3}{2}]} +
T_{[i+n+m+ \frac{3}{2},i+n+m+p+ \frac{5}{2}]}.
\end{equation}
Using that all the renormalized couplings appearing in the right 
hand side in~\eqref{ind-fin} satisfy~\eqref{theo-res}, one obtains 
that 

\barray 
\label{ind-fin1}
T_{[i - \frac{1}{2},i+n+m+p+ \frac{5}{2}]}  & =  &  
\sum\limits_{j=0}^{n}(-1)^j\, T_{i+j} -
\sum\limits_{j=0}^{m}(-1)^j\, T_{i+n+1+j} +
\sum\limits_{j=0}^{p}(-1)^j\, T_{i+ n+m+2+j} \, \\ 
& =  &  
\sum\limits_{j=0}^{n+m+p+2}(-1)^j\, T_{i+j} 
\nonumber 
\earray 
that reproduces Eq.~\eqref{theo-res} for the renormalized coupling of the block. 
This gives the proof of the desired result.

A few comments are in order to show the relevance of our result. First, Eq.~\eqref{theo-res} provides an exact mapping between the SDRG flow of the  couplings $T_i$ and an {\it alternating} random walk. This 
mapping holds true for any distribution of the initial couplings $J_i$. 
However, Eq.~\eqref{theo-res} does not contain any spatial information 
about the SDRG flow. This means that 
from~\eqref{theo-res} it is not straightforward to reconstruct the information 
about the place where the SDRG processes has  occurred. 
This fact represents an obstacle 
to derive from Eq.~\eqref{theo-res} the scaling of correlation functions or of 
the entanglement entropy. Still, we anticipate that this limitation can be 
overcome for the random inhomogeneous XX chain in the large $h$ limit 
(strongly inhomogeneous limit). This happens because the presence 
of the inhomogeneity provides a simple relation between the SDRG 
step $n$ and the distance from the chain center. More precisely, sites far away from the chain center are usually renormalized at later stages along the SDRG procedure. 

Another important consequence of Eq.~\eqref{theo-res} is that, 
given a region containing $n$ spins, Eq.~\eqref{theo-res} 
allows one to derive the distribution of the renormalized couplings  after 
decimating all the spins. Using the random walk framework, 
one obtains that this is the distribution of 
the final position of the walker after $n$ steps. It is 
straightforward to derive this distribution in the limit $n\to\infty$. 
Clearly, the sum of the even and odd sequences in Eq.~\eqref{theo-res} can be treated 
separately. Both are the sum of independent identically distributed exponential 
variables, that follow the gamma distribution. 
By using that for large $n$ the gamma distribution is well approximated by a normal 
distribution, one has that the sum of the even and odd terms in Eq.~\eqref{theo-res} 
are distributed with $(\pi n)^{-1/2}\exp[(x-n/2)^2/n]$. The renormalized 
coupling after $n$ SDRG steps is obtained as the difference between the sum 
of the odd and even sequences in Eq.~\eqref{theo-res}. This is again a normal distribution 
with zero mean and variance $n$, i.e., 
\begin{equation}
\label{ren-dist}
P\left(T=T_{[i - \frac{1}{2},i+n + \frac{1}{2}]}\right)=
\frac{1}{\sqrt{2\pi n}}e^{-T^2/(2n)}. 
\end{equation}
%

\subsection{Entanglement entropy of random singlet states}
\label{sec-ent-sdrg}

In the following sections we will present numerical results for the von Neumann 
entropy in the random inhomogeneous XX chain. The results are obtained 
by using the SDRG method. At the end of the SDRG procedure one obtains 
a valence bond state (VBS), in which all the spins are paired forming 
singlets. For any VBS configuration, the entanglement entropy $S$ 
between a subsystem $A$ and the rest (see Figure~\ref{fig0}) is proportional 
to the number of singlets that are shared between $A$ and its complement 
$\bar A$. It is straightforward to show that for any VBS state, 
the reduced density matrix $\rho_A$ of subsystem $A$ is written as 
\begin{equation}
\label{rdm}
\rho_A=\bigotimes\limits_{i=1}^{n_{A:A}}\rho_{2S}
\bigotimes\limits_{i=1}^{n_{A:\bar{A}}}\rho_{S},
\end{equation}
where $n_{A:A}$ is the number of singlets between spins in $A$ and 
$n_{A:\bar{A}}$. In~\eqref{rdm}, $\rho_S$ and $\rho_{2S}$ are the density 
matrices of a system of one spin and of a singlet, respectively. Specifically,  
$\rho_{2S}$ is defined as 
\begin{equation} 
\label{r2s}
\rho_{2S} =
\frac{1}{2}\begin{pmatrix}
0 & 0   &  0 & 0\\
0 & 1   & -1 & 0\\
0 & -1  &  1 & 0\\
0 & 0   &  0 & 0
\end{pmatrix},
\end{equation}
in the basis $\left|\uparrow\uparrow\right\rangle$, $\left|\uparrow\downarrow
\right\rangle$, $\left|\downarrow\uparrow\right\rangle$, and $\left|\downarrow
\downarrow\right\rangle$. $\rho_{2S}$ has eigenvalues $0,1$. The reduced density 
matrix $\rho_S$ for one of the spins is 
\begin{equation}
\label{r1s}
\rho_S =
\frac{1}{2}\begin{pmatrix}
1  & 0 \\
0 &  1  
\end{pmatrix}.
\end{equation}
Only $\rho_S$ contributes to the von Neumann entropy of subsystem $A$ 
with the rest. This is because $\rho_{2S}$ has only eigenvalues $0,1$. 
This is also physically expected because $\rho_{2S}$ takes into account 
the singlets between spins in $A$. 
The entropy is obtained 
from~\eqref{rdm}, \eqref{r2s}  and~\eqref{r1s}
\begin{equation}
\label{vn-rs}
S=n_{A:\bar{A}}\ln2.
\end{equation}
From~\eqref{vn-rs} one has that the disorder averaged entropy $\langle S\rangle$ 
is proportional to the average number of singlets $\langle n_{A:\bar A}\rangle$ 
shared between $A$ and its complement $\bar{A}$. 

%
\section{Area-law violation in the random inhomogeneous XX chain}
\label{sec-area}

\subsection{Von Neumann entropy: SDRG results}
\label{sec-vn}

%
\begin{figure}[t]
\includegraphics*[width=0.5\linewidth]{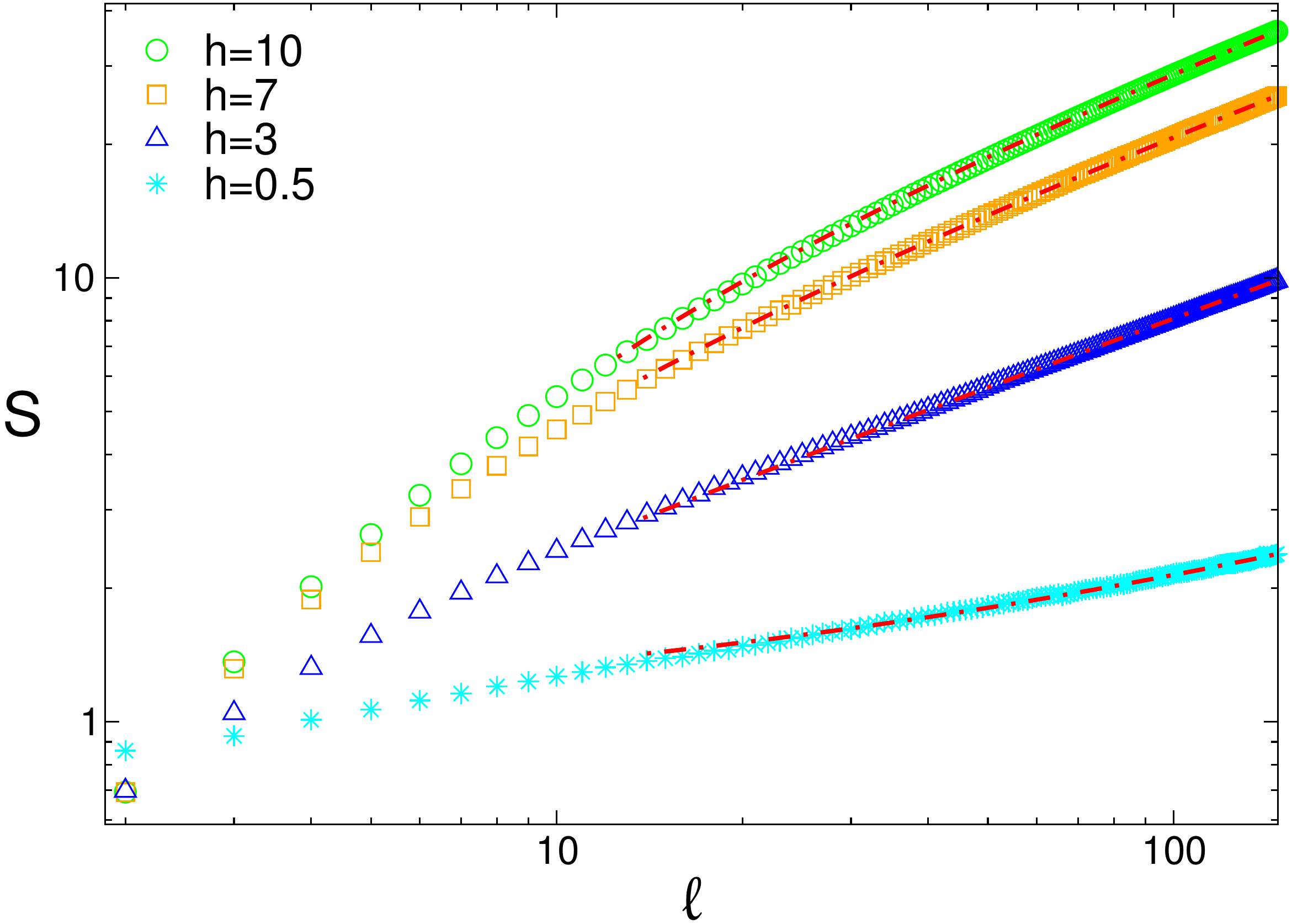}
\caption{Unusual scaling of the entanglement entropy in the randbow phase: 
 Entanglement entropy plotted as a function of the subsystem length $\ell$. 
 The subsystem starts from the center of the chain (see Figure~\ref{fig0}). 
 The data are SDRG results for the random inhomogeneous XX chain. The different 
 symbols are for different values of $h$ and fixed value of $\delta=1$. 
 Logarithmic scale is used on both axes. The dashed-dotted lines are 
 fits to $a+b\ell^{1/2}$, with $a,b$ fitting parameters. 
}
\label{fig-vn}
\end{figure}
%

%
\begin{figure}[t]
\includegraphics*[width=0.5\linewidth]{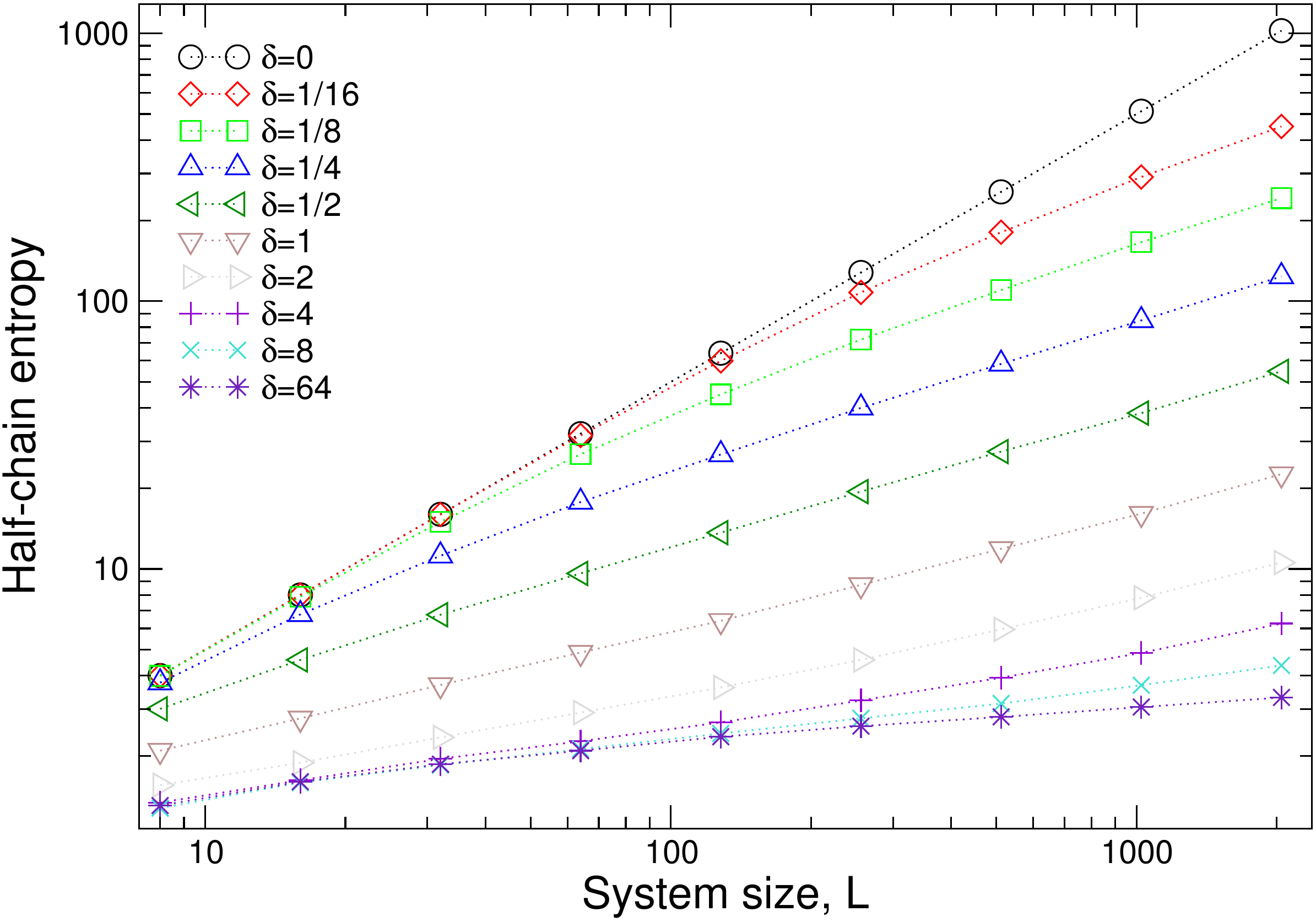}
\caption{Unusual scaling of the entanglement entropy in the randbow phase: 
 Half-chain entanglement entropy  plotted as a function of the chain length $L$. 
 The data are SDRG results for the random inhomogeneous XX chain. The different 
 symbols are for different values of $\delta$ and $h=1$. The data are averaged 
 over $10^4$ disorder realizations. 
}
\label{fig-vn-b}
\end{figure}
%
We now discuss the scaling behavior of the ground-state von Neumann entropy 
in the random inhomogeneous XX chain (cf.~\eqref{ham-XX}). 
In Figure~\ref{fig-vn} we present numerical data for the von Neumann entropy 
$S$ of a subsystem $A$ placed at the center of the chain (see Figure~\ref{fig0}).
The results are obtained by implementing the SDRG method discussed in 
section~\eqref{sec-sdrg}. The entropy $S$ is plotted versus the 
subsystem size $\ell$ of $A$. The different symbols in the Figure correspond 
to different values of the inhomogeneity $h$. The disorder strength parameter 
$\delta$ (cf.~\eqref{coup}) is fixed to $\delta=1$. For $h\to\infty$ the 
model reduces to the rainbow chain, and the volume law $S\propto\ell$ is expected. 
Oppositely, for $h\to0$ the homogeneous random XX chain is recovered with  
logarithmic entanglement scaling~\eqref{rsp-ent}. Surprisingly, for all the 
intermediate values of $0.5<h<10$, the entropy 
exhibits a power-law increase with $\ell$ (notice the logarithmic scale in 
both axes). A preliminary analysis suggests the behavior 
$S\propto\ell^{1/2}$. To perform a more careful finite-size analysis we 
fit the SDRG results to 
\begin{equation}
\label{fit}
S=a+b\ell^{1/2},
\end{equation}
where $a$ and $b$ are fitting parameters. The results of the fits are reported 
in Figure~\ref{fig-vn} as dashed-dotted lines. Clearly, for small values of 
$\ell$ the data exhibit deviations from~\eqref{fit}. This behavior has 
to be attributed to finite-size corrections, due to the small $\ell$. Similar 
corrections are present for clean models, as well as for the random XX 
chain~\cite{laflorencie-2005}. However, already for $\ell\gtrsim 10$ the data are in perfect 
agreement with~\eqref{fit} for all values of $h$ considered. 

%
\begin{figure}[t]
\includegraphics*[width=0.5\linewidth]{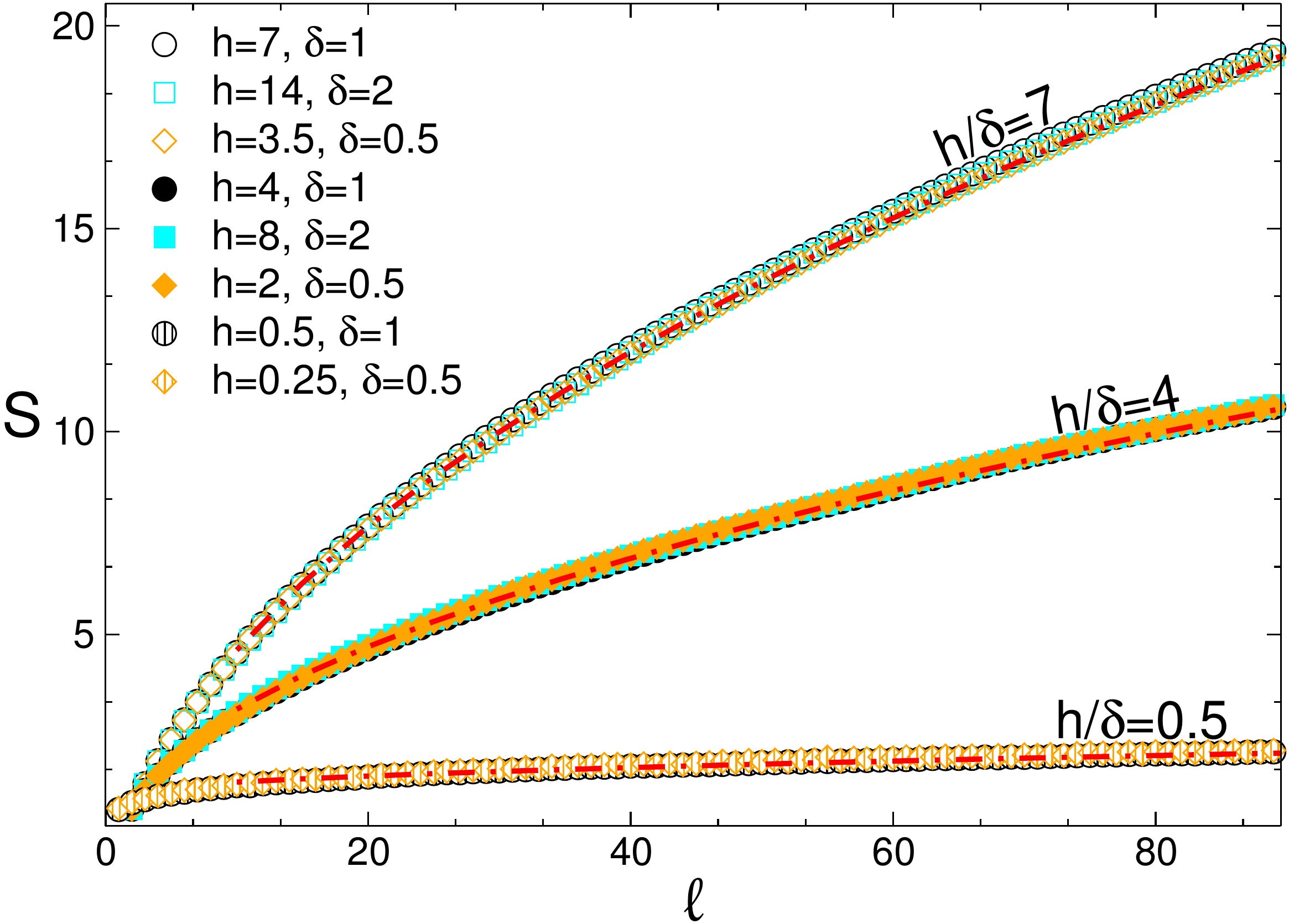}
\caption{Entanglement entropy $S$ plotted as a function of the subsystem size 
 $\ell$: SDRG results for the randbow XX chain. The symbols correspond to 
 several values of $\delta$ and $h$. The data collapse shows that the von Neumann entropy is a function of the ratio 
 $h/\delta$. 
}
\label{fig-vn-a}
\end{figure}
%

Alternatively, in Figure~\ref{fig-vn-b} we plot the half-chain entropy 
as a function of the chain length $L$. The data are now for a wide range 
of $0\le \delta\le 64$. The data are for $h=1$. For $\delta=0$ the volume 
law behavior is visible, whereas the data for $\delta=64$ are suggestive 
of the logarithmic behavior that is expected in the random singlet phase. 
For all other values of $\delta$ the square root scaling is visible, confirming 
the results of Figure~\ref{fig-vn}. 

It is interesting to investigate the combined effect of disorder and 
inhomogeneity on 
the scaling of $S$. This is discussed in Figure~\ref{fig-vn-a}, by considering  
different values of $\delta$ and $h$. The Figure plots $S$ as a function of $\ell$ for 
several values of $h$ and $\delta$ (different symbols). 
All the data for different $\delta$ and $h$ but with the same value of $h/\delta$ 
collapse on the same curve. This confirms  that $S$ is a function of $h/\delta$ 
only, as it was anticipated in section~\ref{sec-sdrg}. 
The Figure shows SDRG results for $h/\delta=7$ (empty symbols), 
$h/\delta=4$ (filled symbols), and $h/\delta=0.5$ (hatched symbols). 
This scaling behavior, however, is valid only within the SDRG method. 
We anticipate that for the random inhomogeneous XX chain the entanglement 
entropy can be calculated exactly (see section~\ref{sec-XX}) using free-fermion 
techniques, and it is a function of $h/\delta$ only for large $h$. 

Finally, for all values of $h,\delta$ considered in Figure~\ref{fig-vn-a} the 
von Neumann entropy exhibits the square-root scaling~\eqref{fit}. 
The dashed-dotted lines in the Figures are fits to~\eqref{fit}, and they are 
in good agreement with the SDRG results. We should also remark that for $h/\delta=0.5$ 
the square root scaling of the von Neumann entropy is visible only for larger 
$\ell\gtrsim 100$, due to larger finite-size effects, as it is also clear also from Figure~\ref{fig-vn}.

\subsection{Understanding the area-law violation: Bubble vs rainbow regions}
\label{sec-bubble}

\begin{figure}[t]
\includegraphics*[width=0.5\linewidth]{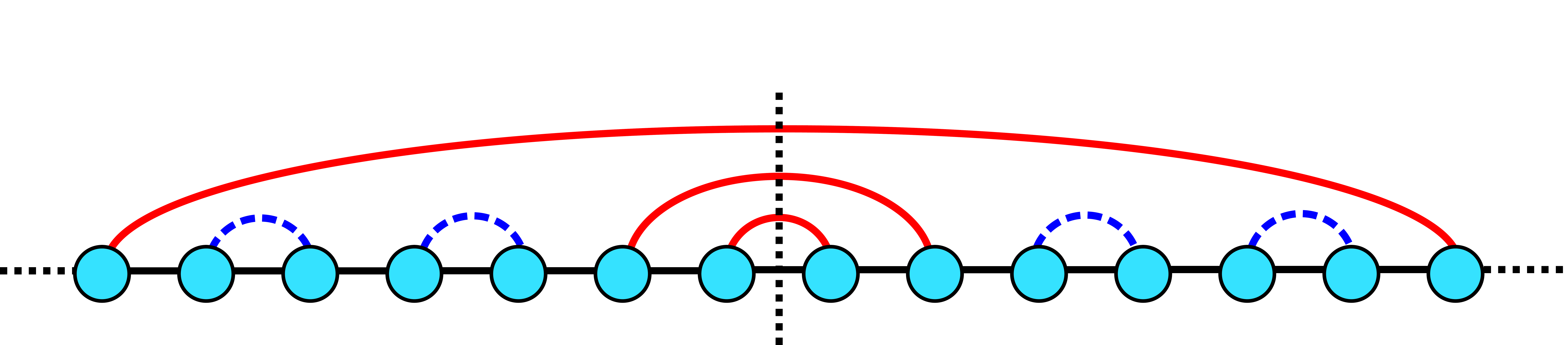}
\caption{ An example of singlet configuration obtained using the 
 SDRG method in the random inhomogeneous XX chain. The limit 
 of strong inhomogeneity $h\gg1$ is considered. The singlet 
 configuration contains two rainbow regions with two and one 
 long singlets, respectively. These are denoted with continuous 
 lines. A bubble region with four short-range singlet denoted by 
 dashed lines is present. 
}
\label{fig-cartoon0}
\end{figure}

The square-root entanglement scaling discussed in Figures~\ref{fig-vn} 
and~\ref{fig-vn-b} can be qualitatively derived from the distribution of the 
rainbow and bubble regions of the states. To this end we shall
define $\ell_r$ as the number of consecutive concentric bonds that constitute a given rainbow region.
For example, in the pure rainbow state we have $\ell_r =L$ bonds connecting the left and right halves of the chain. 
On the other hand, 
we define $\ell_b$ as the number of points that are connected by consecutive
dimer bonds  that constitute a bubble region (see Figure~\ref{fig0-a}).
An  example of a  VBS state for a chain with $2L=14$ sites  is shown in 
Figure~\ref{fig-cartoon0}. This configuration contains  two rainbow regions 
with  $\ell_r=2$ and $\ell_r=1$ bonds  (continuous links), and two  bubble 
regions with  $\ell_b=4$ sites each (see dashed lines in the Figure).

To compute the probability distribution of $\ell_r$ and $\ell_b$ 
we apply the SDRG method to decimate all the spins for a 
set of disorder realizations. 
 
\begin{figure}[t]
\includegraphics*[width=0.5\linewidth]{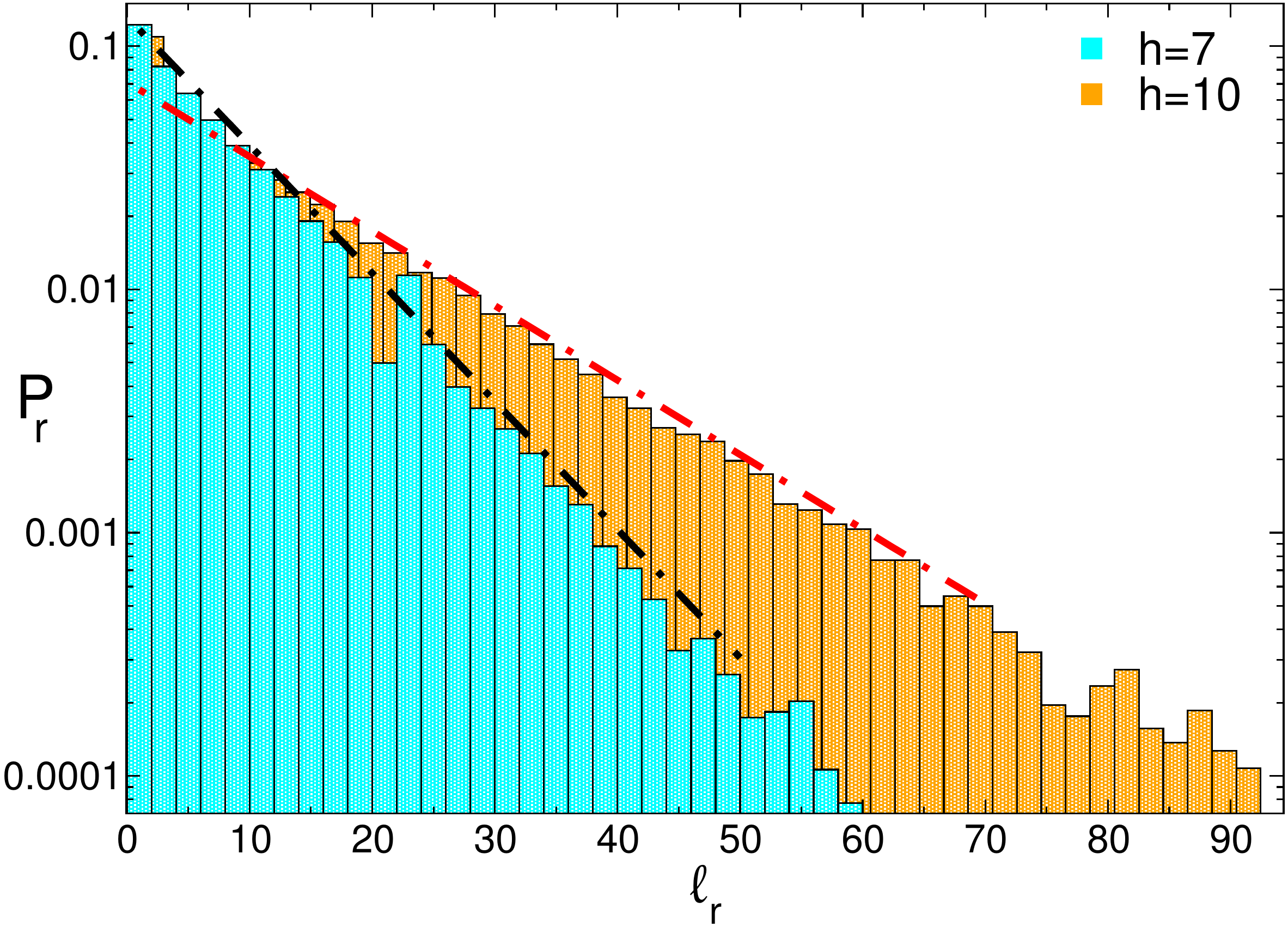}
\caption{Exponential decaying distribution of the rainbow bonds. 
 The probability $P_r(\ell_r)$ of the number of bonds $\ell_r$, of the rainbows. 
 The Figure shows normalized histograms for the distribution of $\ell_r$. 
 The data are SDRG results for the random inhomogeneous XX chain. 
 The histograms correspond to the  values $h=7,10$ and $\delta=1$. Notice the 
 logarithmic scale on the $y$-axis. The dashed-dotted lines are exponential 
 fits. The data are obtained by averaging over $\sim 1000$ different 
 disorder realizations. 
}
\label{fig-Lr}
\end{figure}

The distribution of the rainbow bonds  $P_r(\ell_r)$ is obtained by constructing the 
histograms of the values of $\ell_r$ of the different rainbow regions. 
An average over different disorder realizations is performed. The 
resulting histograms for $\ell_r$ are shown in Figure~\ref{fig-Lr}. 
The data are for the random inhomogeneous XX chain in the strongly 
inhomogeneous limit for $h\gg 1$. 
We use a logarithmic scale on the $y$-axis. The data show a clear 
exponential decay with $\ell_r$. 
The exponential decay is smaller   for 
greater values of  $h$.  
This is an expected result  because in the limit $h\to\infty$ the rainbow regions will start to proliferate.

\begin{figure}[t]
\includegraphics*[width=0.5\linewidth]{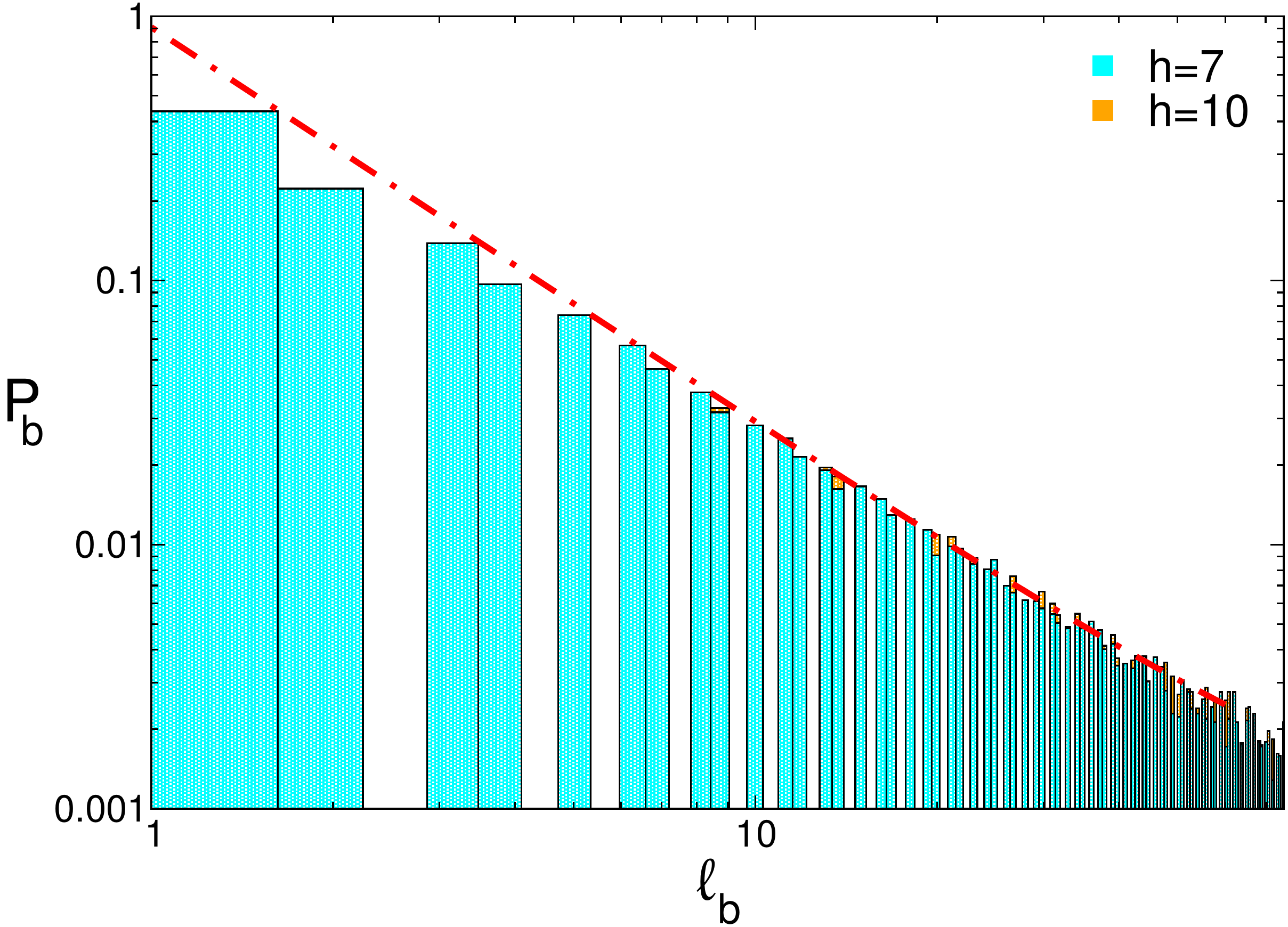}
\caption{Power law distribution of the size of the bubble regions. 
 The Figure shows the probability $P_b(\ell_b)$ of the extension of 
 the bubble regions. Notice the logarithmic scale on both axes.
 The data are renormalized histograms for the length 
 of the bubble phase in the random inhomogeneous XX chain. The data are 
 are obtained using the SDRG method for a chain with $h=7,10$ and 
 $\delta=1$. Each point is obtained by averaging over $\sim 1000$ disorder 
 realizations. The difference between the histograms for $h=7$ and 
 $h=10$ is not visible. The dashed-dotted line is a fit to $\sim x^{-3/2}$. 
}
\label{fig-Lb}
\end{figure}

A similar analysis can be performed for the distribution $P_b(\ell_b)$ of 
the extension $\ell_b$ of the bubble regions. 
The results are reported in Figure~\ref{fig-Lb} for $h=7$ and $h=10$. 
Interestingly, on the scale of the Figure the two histograms are not 
distinguishable, signalling that $P_b$ does not depend significantly on 
$h$, at least for large $h$. In stark contrast with the rainbow regions 
(see Figure~\ref{fig-Lr}), $P_b$ exhibits a power-law decay with $\ell_b$. A 
careful analysis suggests the behavior 
\begin{equation}
\label{b-for}
P_b(\ell_b)\propto\ell^{-3/2}_b. 
\end{equation}
The dash-dotted line in Figure~\ref{fig-Lb} is a fit to the behavior~\eqref{b-for}, 
and it perfectly describes the  numerical data. 

The results of Figure~\ref{fig-Lr} and Figure~\ref{fig-Lb} allow one to 
understand qualitatively   the square-root behavior of the von Neumann entropy~\eqref{sq-ent}. 
First of all, since $P_r$ is an exponential function, 
the average  number  
of rainbow bonds   $\langle\ell_r\rangle$ is a constant independent on $L$, 
\begin{equation}
\langle{\ell}_r\rangle=\int_1^{\infty} dx \,  x P_r(x)  \,  , 
\end{equation}
where we have replaced the upper limit of the integral, namely  $L$ (total number of bonds) by $\infty$,
without changing  essentially the final result. 
On the other hand, given a subsystem $A$ of length $\ell$, 
the  average  number of points  of the bubble regions contained in $A$ is given by 
\begin{equation}
\langle {\ell}_b\rangle=\int_2^\ell dx \,  x P_b(x)\propto\ell^{1/2}.
\end{equation}
The short-range singlets forming the bubble phase do  
not contribute to the entanglement between $A$ and the rest,  
because they mostly entangle spins within $A$. 
The entanglement between $A$ and the rest is due to 
long range links forming the rainbow phase. 
However, the scaling of the entropy is determined by 
the distribution of $\ell_b$, which determines the 
typical spatial separation between the different rainbow 
regions. A crude estimate of the entanglement 
entropy is obtained as follows. On average, there are 
$\langle\ell_r\rangle$ rainbow links every $\langle\ell_b\rangle$ 
sites. Hence a region $A$  with $\ell$ sites can be divided roughly  into 
$\ell/\langle \ell_b \rangle$ bubbles separated by $\langle\ell_r\rangle$  rainbow bonds.
The von Neumann entropy  can then be approximated as 
\begin{equation}
\label{est}
S\propto  \frac{\ell}{\langle\ell_b\rangle} \times  \langle\ell_r\rangle  \ln 2\propto
\ell^{1/2}\langle\ell_r\rangle\ln2,
\end{equation}
i.e., the square-root scaling in Eq.~\eqref{sq-ent}.
Crucially, in~\eqref{est} we have assumed  that the average bubble size $\langle\ell_b\rangle$ 
and average number of  rainbow bonds   $\langle\ell_r\rangle$ 
do not depend on the position in the chain. This might be surprising at first  look 
because the system is not homogeneous. However, as it will be 
clear in the following sections, due to the form of the renormalization 
rule~\eqref{RG-equation} and the type of inhomogeneity, the condition 
that leads to the bubble formation does not depend on the precise SDRG step, 
and, consequently, on the position in the chain. Notice that this relies on 
the precise form of~\eqref{RG-equation}, which holds 
only for the XX model, and it breaks down for the interacting XXZ chain.

%
%

%
%

%
%

\section{Entanglement contour}
\label{sec-cont} 

%
\begin{figure}[t]
\includegraphics*[width=0.55\linewidth]{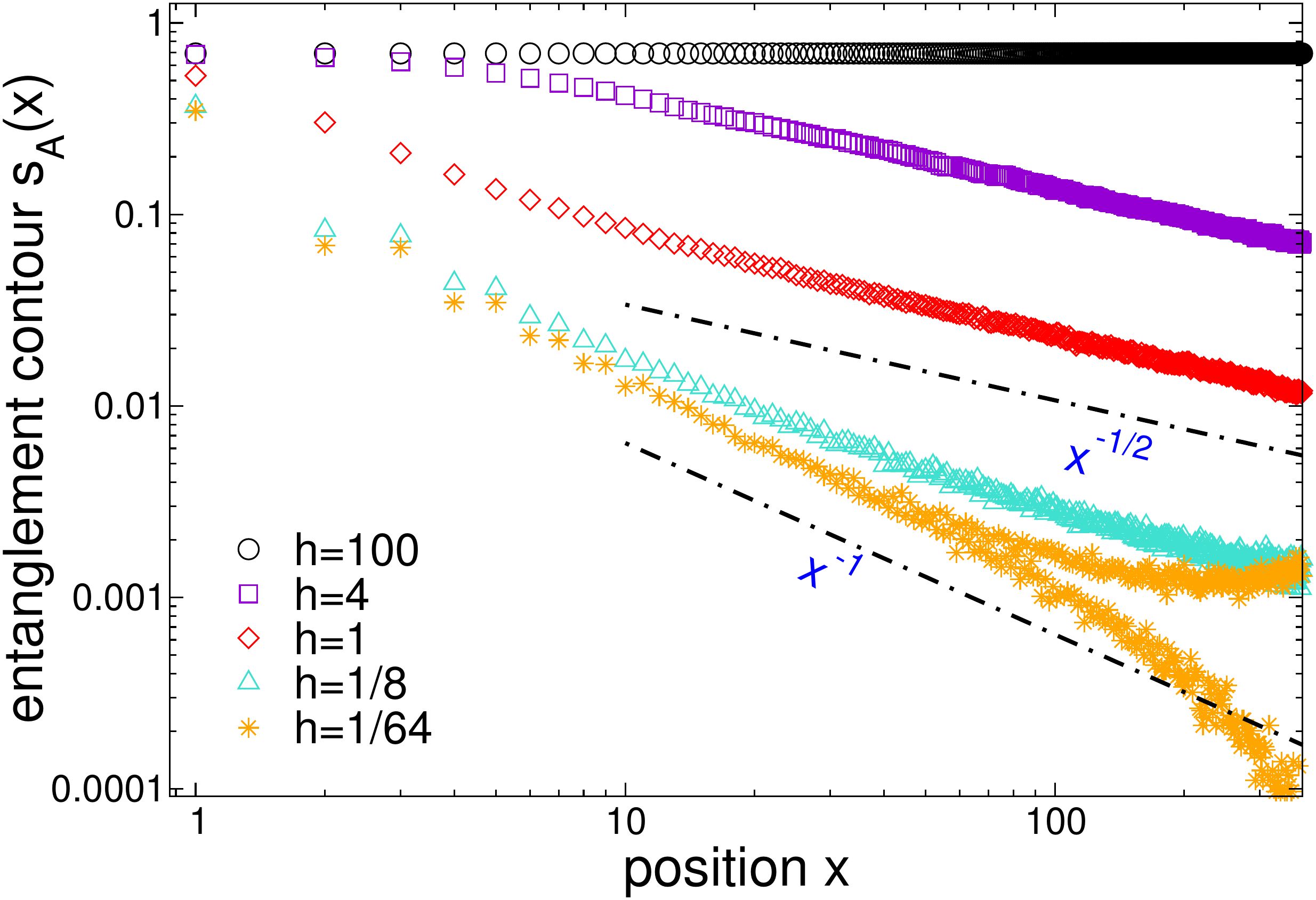}
\caption{ Spatial contributions to the subsystem entanglement entropy $S$ 
 (entanglement contour). The data are SDRG results for the inhomogeneous 
 random XX chain. The Figure shows the contour function $s_A(x)$ 
 as a function of the position $x$ inside block $A$ of the chain. 
 The different symbols correspond to different values of $h$ and 
 fixed $\delta=1$ . For  $h\to\infty$ the contour is flat and equal to 
 $\ln 2$, 
 implying that all the sites of $A$ contribute equally to the 
 von Neumann entropy, which exhibits volume-law scaling. For 
 intermediate values of $0<h<\infty$, one has the 
 behavior $x^{-1/2}$, which implies that $S\propto 
 \sqrt{\ell}$. Finally, in the limit $h\to0$, 
 one has the scaling $s_A\propto x^{-1}$, implying $S\propto\ln\ell$, 
 reflecting the onset of the random singlet phase in that 
 limit. 
}
\label{fig7-aa}
\end{figure}
%

It is enlightening to investigate how the square-root scaling of the von 
Neumann entropy~\eqref{sq-ent} is reflected in the behavior of the 
 entanglement contour. 
The entanglement contour has been introduced in~\onlinecite{vidal-2014} 
as a tool to quantify the spatial contributions to the entanglement 
entropy. The key idea is to write $S$ as the integral of a 
contour function $s_A(x)$, where $x\in A$. The natural constraints that 
$s_A(x)$ has to satisfy are 
\begin{equation}
\label{cont-c}
S=\int_{x\in A} \!\!\!dx s_A(x),\qquad s_A(x)\ge 0. 
\end{equation}
In~\eqref{cont-c} we considered  a continuous system. The extension 
to lattice models is obtained by replacing the integrals with the sum over 
the lattice sites. The first condition in~\eqref{cont-c} is a normalization, 
whereas the second ensures that the contribution of each site to 
the entanglement entropy is positive. Clearly, the conditions~\eqref{cont-c} 
are not sufficient to uniquely identify the contour function $s_A$. 
For instance, for an homogeneous system the simplest choice is the flat 
contour $s_A(x)=S/|A|$, where $|A|$ is the volume of $A$. However, although 
this is a legitimate choice, it does not take into account that, due to the 
area law, most of the contribution to the entanglement between $A$ and its 
complement $\bar{A}$ originates at the boundary between them. For gapped 
one and two dimensional systems this boundary locality of the entanglement 
entropy has been 
thoroughly investigated in Ref.~\onlinecite{b-loc} 
and~\onlinecite{b-loc1}. We should mention that exact calculations of the 
contour function $s_A$ are possible only for free models~\cite{coser-2017,fr-17}, 
and in Conformal Field Theories~\cite{erik} (CFT). 

Within the SDRG framework for random systems, there is a natural definition 
of  entanglement contour. Precisely, the value of $s_A$ on a given site $x$ of 
$A$ is $\ln 2$ if there is a link starting at site $x$ and ending in $\bar{A}$, 
and it is zero otherwise. 
Numerical results for the contour function $s_A(x)$ 
in the random inhomogeneous XX chain as a function of the position $x$ 
in $A$ are shown in Figure~\ref{fig7-aa}. 
We have chosen the block $A$ as
the right half of the chain that contains $L$ sites. 

This  Figure shows the average 
contour function $\langle{s_A}(x)\rangle$ over $\sim 1000$ disorder realizations. 
The position $x=1, 2, \dots, L$ is measured starting from the center of the chain. The 
Figure shows results for several values of $1/64\le h\le 100$. 
Clearly, for $h\to\infty$, the ground state of the model is in the 
rainbow phase (see Figure~\ref{fig0-a} (c)). This implies that $s_A$ 
is flat and $s_A(x)\sim \ln 2$. 

In the limit $h\to0$ the ground state is described by the random singlet 
phase. This is reflected in the behavior of the contour function $s_A$. 
Already for $h=1/64$ the data in Figure~\ref{cont-c} exhibit a 
$\propto1/x$ decay with $x$. 
This implies that for a subsystem of length $\ell$ one has $S=\int_1^\ell s_A(x)
dx\propto\ln\ell$, which is consistent with the expected result~\eqref{rsp-ent}. 
Notice that for $h=1/64$, for large $x$, $s_A(x)$ exhibits large oscillations. 
We do not have an analytic understanding of this behavior, although we should mention 
that similar oscillations in the entanglement were observed for the clean rainbow 
XX chain in Ref.~\onlinecite{erik}. Finally, for intermediate values $1/64<h<100$ 
one has $s_A(x)\propto x^{-1/2}$. This is clearly consistent 
with the square-root scaling behavior $S\propto\ell^{1/2}$, as it was shown in 
Figure~\ref{fig-vn}. Again, this behavior of the contour is a consequence 
of the power-law scaling of the distribution of $\ell_b$. Indeed, the probability 
that a site at distance $x= \ell$ from the chain center contributes to the 
entanglement is roughly $1/\langle\ell_b\rangle\sim 1/\ell^{1/2}$.

\section{Numerical benchmarks using the exact solution of the XX chain}
\label{sec-XX}

%
\begin{figure}[t]
\includegraphics*[width=0.5\linewidth]{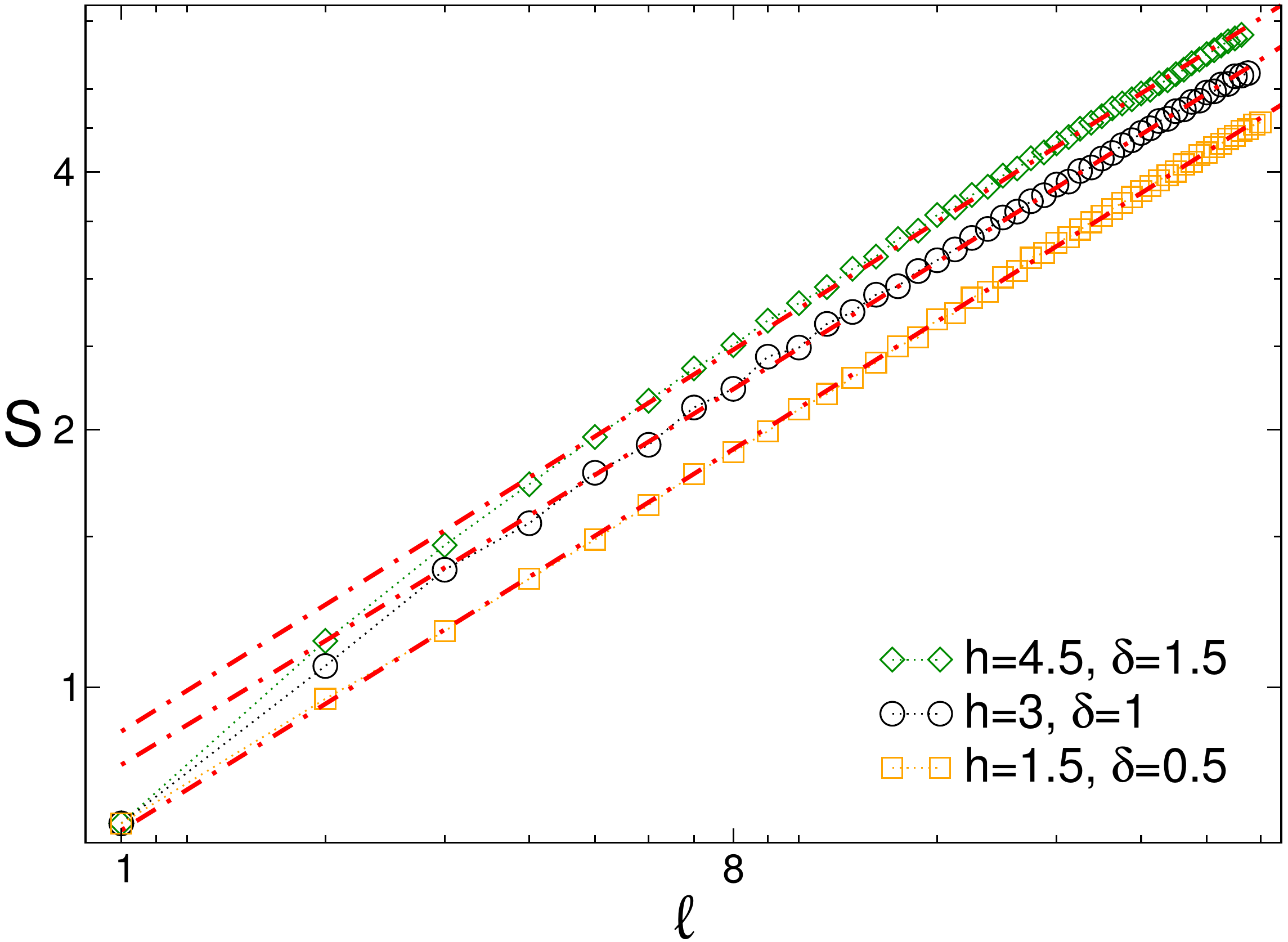}
\caption{Entanglement entropy $S$ in the random inhomogeneous XX 
 chain: Exact results. The Figure shows $S$ obtained using free-fermion 
 techniques plotted versus the size $\ell$ of the subsystem. Note the 
 logarithmic scale on both axes. The symbols are the data for a chain 
 with 
 $2L=100$ sites and several values of $h$ and $\delta$. Each point  
 is obtained by averaging $S$ over $500$ different disorder realizations. 
 The dashed-dotted line is a fit to the expected behavior $S=a+
 b\ell^{1/2}$, with $a,b$ fitting parameters. Notice that $S$ is not 
 a function of the ratio $h/\delta$ only. 
}
\label{fig-XX}
\end{figure}
%

In this section we provide exact 
results for the entanglement entropy of the random inhomogeneous 
XX chain~\eqref{ham-XX}. The key observation is that for any 
disorder distribution, the XX chain is exactly solvable after 
mapping it to free fermions. 
The single-particle eigenstates $|\Psi_q\rangle$ (with $q$ an integer 
that labels the different eigenstates) of~\eqref{ham} are of the form 
\begin{equation}
\eta_q^\dagger|0\rangle\equiv|\Psi_q\rangle=\sum_{i= - L + \frac{1}{2}}^{L - \frac{1}{2}} \Phi_q(i)c_i^\dagger|0\rangle,
\end{equation}
%
%
with $|0\rangle$ denoting the fermionic vacuum, and $\Phi_q(i)$ the eigenstate 
amplitudes. Here $\eta_q$ denotes a new fermionic operator creating the 
single particle excitation. To determine $\Phi_q(i)$ one has to solve the 
Sch\"odinger equation, which reads 
\begin{equation}
\label{xx-eig}
- J_{i+ \frac{1}{2}}  \Phi_q(i+1) - J_{i- \frac{1}{2}}\Phi_q(i-1)= 2 \epsilon_q\Phi_q(i),\quad i= \pm \frac{1}{2}, \pm \frac{3}{2}, \dots, \pm (L - \frac{1}{2} ), 
\end{equation} 
with $J_L= J_{-L} =  0$, and $\epsilon_q$ the single-particle energies. Eq.~\eqref{xx-eig} 
defines the eigenvalue problem for the banded  $(2L) \times (2L)$  matrix 
$T_{i,j} \equiv   \frac{1}{2} (J_{j+ \frac{1}{2}} \delta_{i,j+1}+J_{j- \frac{1}{2}}\delta_{i,j-1})$. 
The eigenvalues of $T$ are 
organized in pairs with opposite sign. This can be shown as follows. 
Given the amplitude $\Phi_1(i)$ of an eigenvector with $\epsilon_q
>0$, it is straightforward to check that the amplitudes of the eigenvector 
with eigenvalue $-\epsilon_q$ are obtained as $(-1)^{i}\Phi_q(i)$. 
The ground state $|GS\rangle$ of~\eqref{ham} at half filling 
 belongs to  the sector with  
$M = L$
fermions, and it is constructed by filling all the negative modes as  
\begin{equation}
|GS\rangle=\eta^\dagger_{q_M}\eta^\dagger_{q_{M-1}}\cdots\eta_{
q_1}^\dagger|0\rangle.
\end{equation}
It is useful for the following to derive the anticommutation relations
\begin{equation}
\label{us1}
\{\eta^\dagger_q,c^\dagger_j\}=\{\eta_q,c_j\}=0, 
\end{equation}
and
\begin{equation}
\label{us2}
\{\eta_q^\dagger,c_j\}=\Phi_q(j)\delta_{k,j},\quad\{\eta_q,
c^\dagger_j\}=\Phi^*(j)\delta_{k,j}. 
\end{equation}
Using~\eqref{us1} and~\eqref{us2}, the expectation value of the 
two-point function $\langle c^\dagger_ic_i\rangle$ in a generic 
eigenstate of~\eqref{ham-XX} reads 
\begin{equation}
\label{corr-f}
\langle c_i^\dagger c_j\rangle=\sum_{q}\Phi_q^*(i)\Phi_q(j), 
\end{equation}
where the sum if over the filled modes $q$ defining the eigenstate.

Let us now consider the bipartition of the chain in Figure~\ref{fig0}. For any 
free-fermion model, even in the presence of disorder, the reduced density matrix 
$\rho_A$ of subsystem $A$ can be obtained from the correlation matrix~\cite{peschel-1999,
peschel-1999a,chung-2001,peschel-2004,peschel-2004a,peschel-2009} restricted to 
$A$ 
\begin{equation}
{\mathcal C}^{(A)}_{ij}\equiv\langle c_i^\dagger c_j\rangle, 
\end{equation}
where $i,j\in A$. Moreover, given the eigenvalues $\lambda_k$ of 
${\mathcal C}^{{(A)}}$, the entanglement entropy $S$ is given as 
\begin{equation}
\label{ent-ff}
S=-\sum_k(\lambda_k\ln\lambda_k+(1-\lambda_k)\ln(1-\lambda_k)).
\end{equation}
Numerical results for the von Neumann entropy $S$ obtained using~\eqref{ent-ff} 
are reported in Figure~\ref{fig-XX} versus the subsystem size $\ell$. 
The Figure shows results for a chain with 
$2L=100$  sites and several values of 
$h$ and $\delta$ (different symbols in the Figure). We should mention 
that due to the exponential decay of the couplings $J_i$, the calculation of the 
eigenvalues $\lambda_j$ requires to use arbitrary precision routines. The 
results in Figure~\ref{fig-XX} were obtained 
requiring precision up to $10^{-80}$. To highlight the power-law behavior of 
$S$, in the Figure we use a logarithmic scale on both axes. Clearly, for all values of 
$h$ and $\delta$, the data exhibit the behavior $S\propto
\ell^{1/2}$. The dashed-dotted lines in the Figure are fits to  
\begin{equation}
S=a+b\ell^{1/2}, 
\end{equation}
with $a,b$ fitting parameters. For small 
values of $h$ the asymptotic scaling of $S$ is already visible for 
$\ell\gtrsim3$, whereas upon increasing $h$ the asymptotic scaling 
sets in at larger values of $\ell$, as expected. We should also mention that 
the finite-size effects due to $L$ are negligible. This 
is expected because the subsystem is placed at the center of the chain. 
One should observe that all the data shown in the Figure correspond to the 
same value of $h/\delta=3$.  Surprisingly, no data collapse is 
observed, suggesting that the entropy is not a function of the 
ratio $h/\delta$ only. This is in contrast with the SDRG data 
(see Figure~\ref{fig-vn-a}), for which the scaling with $h/\delta$ 
holds.

\section{A toy model for the strongly inhomogeneous limit}
\label{sec-toy}

%
\begin{figure}[t]
\includegraphics*[width=0.75\linewidth]{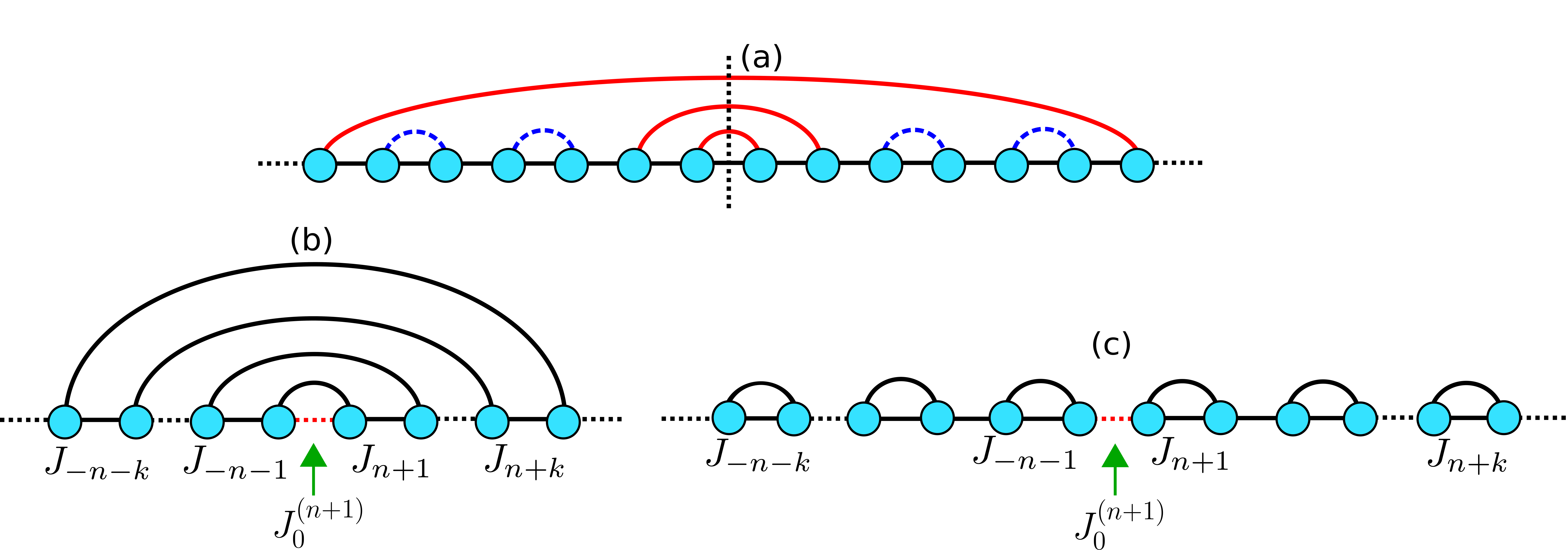}
\caption{ (a) Typical singlet configuration in the ground state of the 
 random inhomogeneous chain in the limit $h\gg1$. The continuous 
 lines denote long-range singlets (``rainbow'' configurations), 
 whereas the dashed ones are short-range singlets connecting 
 spins on nearest-neighbor sites and 
 forming the ``bubble'' phase. Notice that only symmetric link configurations 
 with respect to the center of the chain (marked by the vertical line) 
 are allowed. (b) A rainbow diagram formed by long links connecting 
 distant spins across the chain center. The renormalized central link 
 $J_0^{\scriptscriptstyle(n+1)}$ results from link configurations as 
 in (a). The length of the diagram is denoted as $k$. (c) Bubble 
 diagram of length $k$ formed by short singlets joining spins on 
 nearest-neighbor sites. Notice in both (b) and (c) the symmetry with 
 respect to the chain center. In the limit $h\to\infty$ typical bond 
 configurations as in (a) are obtained by combining rainbow and bubble 
 diagrams ((b) and (c)). 
}
\label{fig1}
\end{figure}
%
In this section we discuss the strongly inhomogeneous limit of the random 
XX chain (cf.~\eqref{ham-XX}), which is obtained for $h\to\infty$ 
in~\eqref{coup}. In this limit, several analytical results can be obtained, 
for instance the scaling of the survival probabilities for the rainbow and 
the bubble regions presented in Figure~\ref{fig-Lr} and Figure~\ref{fig-Lb}. 
For $h\gg1$, the ground state of~\eqref{ham-XX} has the structure presented 
in Figure~\ref{fig0-a} (b). This consists of long links forming a ``rainbow'' phase connecting  
distant spins across the chain center, and of short links connecting spins 
on neighboring sites, forming a  ``bubble'' phase. Importantly, for large 
$h$ all the link configurations are symmetric with respect to the center 
of the chain. This is due to the fact that for large $h$, 
more 
degrees of freedom (bonds), which are decimated first, which 
are nearer   
to the  center of the chain. This implies that SDRG decimations happen 
symmetrically with respect to the chain center. To further enforce 
this symmetry in the following we will restrict ourselves to symmetric 
couplings, i.e., $K_n=K_{-n}$  (cf.~\eqref{coup}). 
A  crucial consequence of the large $h$ limit is that the net effect of the 
SDRG procedure, at any step, is to renormalize the central coupling $J_0$ 
(see~\eqref{ham}). Moreover, the VBS state obtained at the end of the 
renormalization is constructed using only two types of diagrams that 
we term ``rainbow diagrams'' and ``bubble diagrams''. A typical 
singlet configuration is depicted in Figure~\ref{fig1} (a). 
The building blocks, i.e., rainbow and bubble diagrams, 
are better discussed in Figure~\ref{fig1} 
(b) and Figure~\ref{fig1} (c), respectively. In both    (b) and (c) 
the coupling $J_0^{\scriptscriptstyle (n+1)}$, connecting sites $n+1$ and 
$-n-1$, is the renormalized coupling obtained after decimating the first $2n$ 
spins around the chain center 
(in this section the position of the spins are labelled
by integers: $\pm 1, \pm 2, \dots, \pm L$).  
In the following we derive exact analytic expressions for 
$J^{\scriptscriptstyle(n+1)}_0$. Also, by using~\eqref{theo-res} 
we establish a relation between the survival probability of the rainbow and 
bubble diagrams with certain survival probabilities of an  alternating random walk.

\subsection{Rainbow diagrams: Random walk \& survival probability}
\label{r-diag}
%

%

Here we discuss the renormalized coupling obtained from the rainbow 
configuration illustrated in Figure~\ref{fig1} (b). First, the initial  
coupling $J_0^{\scriptscriptstyle(n+1)}$ can result from 
both rainbow and bubble configurations. 
We now consider the effect of a rainbow diagram of $k$ links.  
This is obtained by decimating $k+1$ spin pairs around 
the chain center. Using the strong disorder RG rule~\eqref{RG-equation} 
the renormalized coupling $J_0^{\scriptscriptstyle(n+k+1)}$ that connects 
the spins at sites $n+k+1$ and $-n-k-1$ is given as 
\begin{equation}
\label{rain_coup}
J_0^{(n+k+1)}= [J_0^{(n+1)}]^{(-1)^{k}}\Big[\prod\limits_{\alpha=0}^{k-1}
\Big(J_{n+\alpha+1}J_{-(n+\alpha+1)}\Big)^{(-1)^\alpha}\Big]^{(-1)^{k-1}}.
\end{equation}
As in Eq.~\eqref{theo-res}, 
it is useful to take the logarithm 
of~\eqref{rain_coup} obtaining 
\begin{equation}
\label{rain_coup_log}
-\ln J_0^{(n+k+1)}=(-1)^{k-1}\Big[-X_0^{(n+1)}+\sum
\limits_{\alpha=0}^{k-1}(-1)^\alpha(X_{n+\alpha+1}+X_{-n-\alpha-1})
\Big]+(n+k+ 1/2 )h. 
\end{equation}
Here we defined $X_j\equiv -\ln K_j$. In~\eqref{rain_coup_log}, the term 
$(n+k+1/2)h$ is the contribution of the inhomogeneity (cf.~\eqref{coup}). 
Importantly, in~\eqref{rain_coup_log}, $X_0^{\scriptscriptstyle(n+1)}$ is 
obtained from $-\ln J_0^{\scriptscriptstyle (n+1)}$ by considering only the 
contributions of $K_j$ (cf.~\eqref{coup}), i.e., it does not take into account 
the contribution of $h$, which is included in the last 
term in~\eqref{rain_coup_log}. Crucially, here we are using that the 
$h$-dependent term in~\eqref{rain_coup_log} does not depend on 
the renormalization pattern leading to $J_0^{\scriptscriptstyle(n+1)}$.  
This is a simple consequence of~\eqref{theo-res}. Specifically, we 
observe that the $h$-dependent term in 
$-\ln J_0^{\scriptscriptstyle(n)}$ is $(n-1/2)h$, from which the last term 
in~\eqref{rain_coup_log} follows. 
This is easy to prove by induction. The 
proof is a simpler version of that for~\eqref{theo-res}. One first assumes  
that after decimating all the spins between sites $n$ and $-n$ the 
$h$-dependent contribution to the coupling is given 
by $(n+1/2)h$. Then one considers the two possible 
SDRG processes, which consist in adding a rainbow link between the spins 
at $(n+1)$ and $-(n+1)$, or two short links connecting spins $(n+1)$ and $(n+2)$ and 
the spins $-(n+1)$ and $-(n+2)$, respectively. It is trivial to verify 
that in both cases the formula holds. 

%
\begin{figure}[t]
\includegraphics*[width=.7\linewidth]{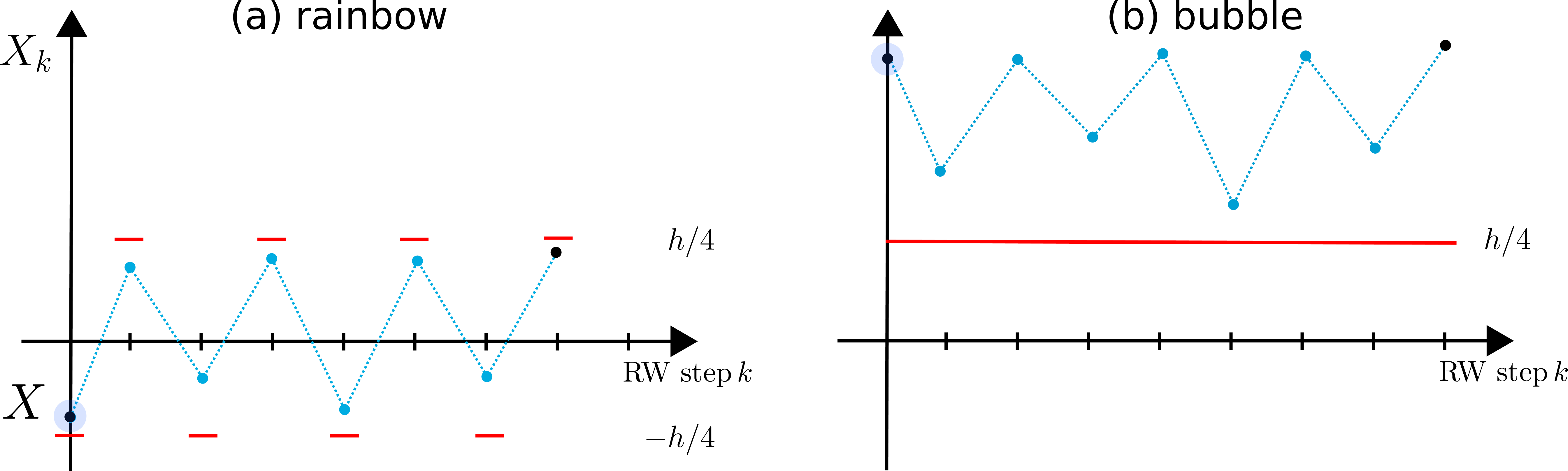}
\caption{ Random walk interpretation of the rainbow and bubble diagrams 
 (see Figure~\ref{fig1} (b) and (c), respectively). (a) The probability 
 for a rainbow diagram to survive $k$ SDRG steps is mapped to the probability 
 for an {\it alternating} random walk to be confined in the alternating strip 
 between $[-h/4,h/4]$ for $k$ consecutive steps. (b) 
 The probability for a bubble phase 
 to survive $k$ SDRG steps is the probability for the random walk 
 to stay above the line $h/4$ for $k$ consecutive steps. In both (a) and (b) 
 the initial point of the walker $x$ is related to the renormalized central 
 bond $J_0$ in Figure~\ref{fig1} (b) and (c). 
}
\label{fig5}
\end{figure}
%
It is important to observe in~\eqref{rain_coup_log} the overall alternating term $(-1)^{k-1}$ 
and the alternating term $(-1)^\alpha$. We anticipate 
that the former is crucial to determine the survival probability of the rainbow 
diagrams. 
Here we are interested in the probability that the rainbow diagram survives 
$k$ successive SDRG 
decimation steps. Crucially, while this survival probability could depend on the 
history of SDRG process, for the XX chain this is not the case, as we are going to show. 
Given a rainbow diagram of length $k$, we start by  calculating the probability for 
the diagram to survive for an extra SDRG step. 
In terms of the couplings $J_i$ (cf.~\eqref{coup}), the survival condition is
\begin{equation} \label{condition-rainbow}
J_0^{(n+k+1)}>J_{n+k+1}, 
\end{equation}
which ensures that an extra rainbow link is created by decimating the spins 
at positions $-n-k-1$ and $n+k+1$. 
Equivalently, in terms of the logarithmic variables $X_k$ 
(cf.~\eqref{rain_coup_log}) Eq.~\eqref{condition-rainbow} reads
\begin{equation}
\label{rw-R1}
(-1)^{k-1}\Big[X+\sum\limits_{\alpha=0}^{k-1}(-1)^\alpha
X_{n+\alpha+1}\Big]<\frac{h}{4}+\frac{1}{2}X_{n+k+1}, 
\end{equation}
where we used that $X_{\alpha}=X_{-\alpha}$ and we defined $X=-X_0^{(n+1)}/2$ 
as the starting point of the random walk.  The survival probability 
condition~\eqref{rw-R1} does not depend on $n$ and $k$. The 
linear term in $n+k$ in~\eqref{rain_coup_log} cancels out with the $h$ dependent 
term in $J_{n+k+1}$. We anticipate that this is not the case in presence of interactions, 
i.e., for the XXZ chain, and it will have striking consequences for the scaling of the 
von Neumann entropy (see section~\ref{sec-int}). To further simplify the condition~\eqref{rw-R1}, in the 
following we  shall  neglect the term $X_{n+k+1}$. For large enough $h$ this 
should be allowed because $X_{n+k+1}$ is exponentially distributed in 
$[0,\infty]$. The condition in Eq.~\eqref{rw-R1} has a simple interpretation  in terms of random walks.
Due to the factor $(-1)^{k-1}$ the rainbow 
survival probability is the probability of a walker to stay below $h/4$ if 
$(k-1)$ is even and above $-h/4$ if $(k-1)$ is odd, remaining 
confined in the alternating strip $[-h/4,h/4]$. This 
is illustrated pictorially in Figure~\ref{fig5} (a). Interestingly, 
the probability that the walker survives within the strip for $n$ steps decays 
{\it exponentially} with $n$. The details of the calculation are reported 
in Appendix~\ref{sec-calc}.

\subsection{Bubble diagrams: Random walk \& survival probability} 
\label{b-diag}

We now discuss the survival probability for the bubble diagram. 
The typical bubble diagram is shown in Figure~\ref{fig1} (c), 
and it consists of a sequence of short-range singlets between nearest 
neighbor spins. Here we restrict 
ourselves to the situation in which the bubble diagrams  appear in pairs 
(i.e., symmetrically) around the chain center, which is a consequence of the 
choice $J_m=J_{-m}$. Similar to the rainbow diagrams, the net effect of 
bubble diagrams is to renormalize the central coupling $J_0^{\scriptscriptstyle(n+1)}$. 
After a repeated application of the SDRG rule~\eqref{RG-equation}, 
the renormalized coupling $J_0^{\scriptscriptstyle(n+1+2k)}$ for the diagram 
in Figure~\ref{fig1} (c) is given as 
\begin{equation}
\label{b-coup}
J_0^{(n+1+ 2k)}=J_0^{(n+1)}\prod\limits_{\alpha=0}^{2k-1}\Big[
J_{n+\alpha+1}J_{-n-\alpha-1}\Big]^{(-1)^{\alpha+1}}.  
\end{equation}
It is convenient to use logarithmic variables to obtain 
\begin{equation}
\label{sin_coup_log}
-\ln J_0^{(n+1+2k)}=X_0^{(n+1)}+\sum\limits_{\alpha=0}^{2k-1}(-1)^{\alpha-1}
(X_{n+\alpha+1}+X_{-n-\alpha-1})+(n+2k+1/2)h. 
\end{equation}
Again, the flow of the renormalized coupling 
in~\eqref{sin_coup_log} can be interpreted as  a random walk with 
starting point $X_0^{\scriptscriptstyle(n+1)}$. In contrast  to 
the rainbow, there is no overall oscillating term $(-1)^{k-1}$, and 
the walker can only make an even number of steps, because bubbles are 
produced in pairs. The last 
term in~\eqref{sin_coup_log} encodes the inhomogeneity contribution 
to the renormalized coupling, and it is independent on the renormalization 
pattern, as for the rainbow diagram. The condition for the bubble 
diagram to survive two SDRG steps is 
\begin{equation} \label{condition-bubble}
J_0^{(n+1+2k)}<J_{n+1+2k}. 
\end{equation}
In the logarithmic variables one finds 
\begin{equation} \label{rw-B}
X+\sum\limits_{\alpha=0}^{2k-1}(-1)^{\alpha-1}
X_{n+\alpha+1} > \frac{h}{4}+\frac{1}{2}X_{n+2k+1}. 
\end{equation}
where now the starting point of the random walk is defined 
as $X=X_0^{\scriptscriptstyle(n+1)}/2$. Similar to the rainbow, in the 
following we neglect the term $X_{n+2k+1}$ in~\eqref{rw-B}, because it does 
not affect the qualitative behavior of the results. 
In the random walk language, the condition~\eqref{rw-B} defines  
the probability that the walker stays above the line $h/4$, as 
depicted in Figure~\ref{fig5} (b). Importantly, the survival condition 
does not depend on the SDRG step, due to the cancellation of the linear 
term in $n$ in~\eqref{condition-bubble}. Now, the probability that 
the walker satisfies~\eqref{rw-B} for $n$ steps decays as $n^{-3/2}$, 
in contrast with the rainbow survival probability, which decays 
exponentially. This is a standard calculation in the random walk 
literature. We report the details in Appendix~\ref{sec-calc}. 


%
\begin{figure}[t]
\includegraphics*[width=0.5\linewidth]{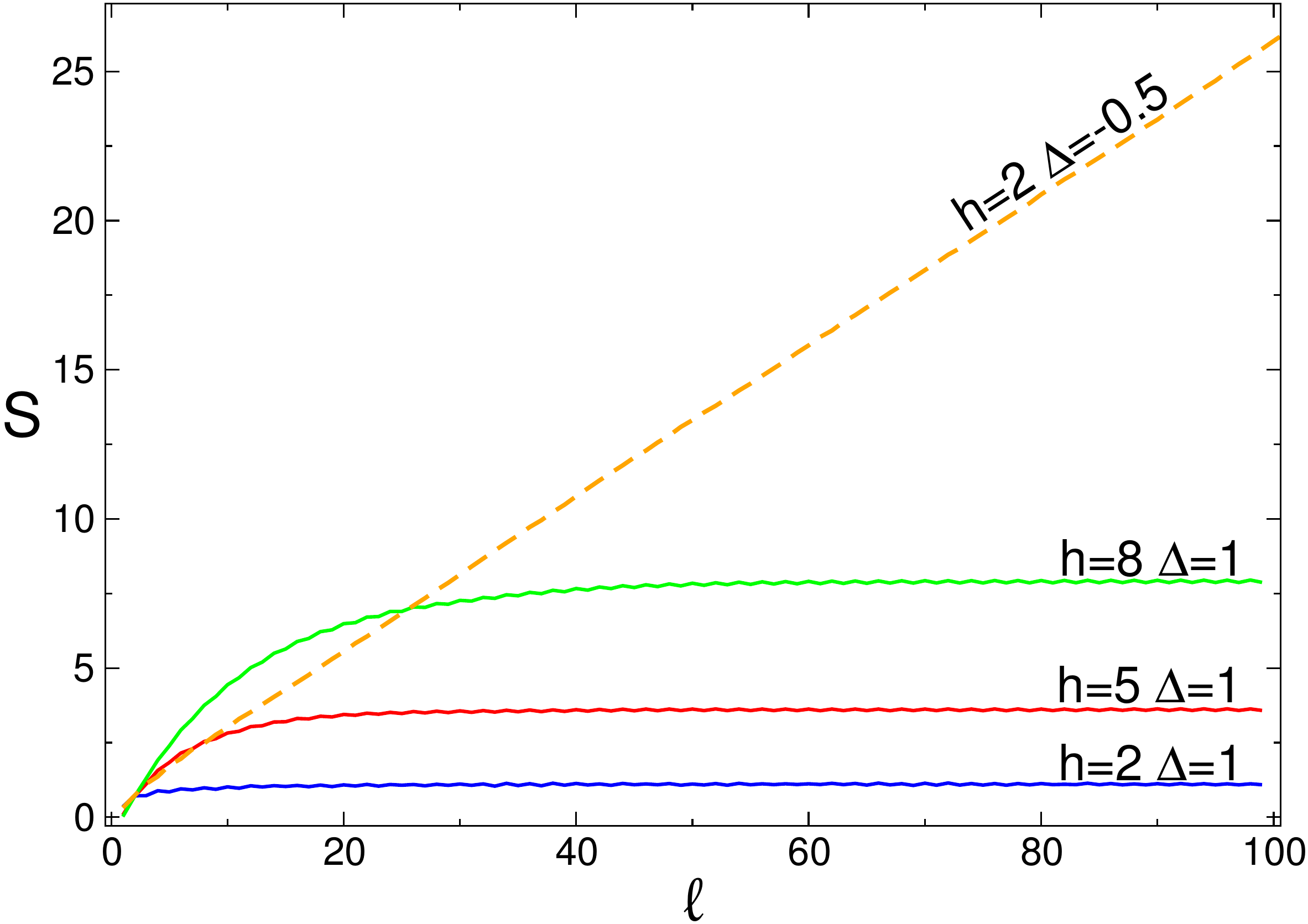}
\caption{Scaling of the entanglement entropy in an interacting inhomogeneous 
 model. The Figure shows the von Neumann entropy $S$ as a function of 
 $\ell$, $\ell$ being the subsystem size. The curves are SDRG results 
 for the randbow XXZ chain for different values of $h$ and $\Delta$. 
 In contrast with the XX chain, $S$ exhibits a saturating behavior for 
 $\Delta>0$, whereas the volume-law behavior $S\propto\ell$ is observed for 
 $\Delta<0$. In particular, the dashed-dotted line is the SDRG result for 
 $\Delta=-0.5$ and $h=2$. 
}
\label{entanglement-int}
\end{figure}
%

\section{Entanglement entropy in the interacting case}
\label{sec-int}

Having established that in the random inhomogeneous XX chain the entanglement 
entropy exhibits an unusual area-law violation, it is natural to investigate 
whether this scenario survives in the presence of interactions. 
In this section we show that the square-root scaling of the entropy~\eqref{sq-ent} 
is very fragile, and it does not survive if the model is interacting. 
To be specific, here we consider the inhomogeneous random Heisenberg XXZ 
chain. This is defined by the hamiltonian
\begin{equation}
\label{h-xxz}
H=\sum\limits_{i=1}^{L-1}J_i\Big\{\frac{1}{2}\Big[S_i^+S_{i+1}^-
+S_i^-S_{i+1}^+\Big]+\Delta S_i^zS_{i+1}^z\Big\}. 
\end{equation}
Here $\Delta$ is an anisotropy parameter. The  XXX chain corresponds 
to the isotropic limit $\Delta=1$. In~\eqref{h-xxz}, $J_i$ 
are the same as in~\eqref{coup}. For simplicity, we choose $J_i=J_{-i}$. 
The SDRG method for the Heisenberg chain is similar to that for the 
XX chain. The only difference (see~\eqref{RG-equation}) is a factor $1+\Delta$ 
in the coupling renormalization~\cite{igloi-rev}. Precisely, the SDRG rule for 
the renormalized coupling $J'$ in the XXZ chain now reads~\cite{hoyos-2007} 
\begin{equation}
\label{rc-1}
J'=\frac{J_LJ_R}{(1+\Delta)J_M}, 
\end{equation}
where, as in~\eqref{RG-equation}, $J_M$ is the largest coupling. In 
the random {\it homogeneous} XXZ chain (i.e., for $h=0$) the 
factor $1+\Delta$ in~\eqref{rc-1} is irrelevant in the scaling limit of large 
systems. 
For instance, the SDRG fixed point describing the ground state 
is the same for both the XX and the XXX chain. This implies that universal 
properties are the same for both models. The entanglement entropy 
exhibits the logarithmic growth~\eqref{rsp-ent}, and the prefactor 
of the logarithm does not depend on $\Delta$.

The goal of this section is to show that in the presence of inhomogeneous 
couplings the factor $1+\Delta$ in~\eqref{rc-1} dramatically changes 
this picture, at least within the framework of the SDRG method. 
The results are discussed in 
Figure~\ref{entanglement-int}. The Figure shows SDRG data 
for the von Neumann entropy $S$ of a subsystem at the center of 
the chain plotted as a function of the subsystem size 
$\ell$. The continuous lines in the figure correspond to several values 
of the inhomogeneity parameter $h$ and $\Delta=1$. 
Surprisingly, for all values of $h$, $S$ saturates in the limit $\ell\to\infty$. 
For $h=8$ there is a large intermediate region where 
the square-root scaling behavior~\eqref{sq-ent} holds. This signals the presence 
of an $h$ dependent crossover length scale $\xi_h$ separating the square-root 
behavior from the saturating behavior at $\ell\to\infty$. 
This behavior changes dramatically for $\Delta<0$. For instance, the dashed 
line in Figure~\ref{entanglement-int} denotes the SDRG data for $h=2$ and 
$\Delta=-1/2$. Clearly, the entanglement entropy exhibits the volume-law 
scaling $S\propto\ell$. We numerically observed that this volume-law scaling 
happens generically for $\Delta<0$.

\subsection{Random walk interpretation}
\label{sec-int1}

We now discuss the origin of the behavior observed in Figure~\ref{entanglement-int}. 
Here we focus on the limit $h\gg1$, where one can exploit the mapping 
between the SDRG flow and the alternating random walk. 
First, in the large $h$ limit, similar to the non-interacting case, 
higher-energy degrees of freedom are nearer to the chain center, and 
are decimated first. This implies that 
the effect of the SDRG procedure is to renormalize the central 
coupling, similar to the XX chain. It is also natural to expect that 
for large $h$ the most likely SDRG patterns are the rainbow and bubble 
patterns discussed in Figure~\ref{fig1} (b) and (c).

To proceed, we first discuss the renormalization of $J_0$ 
due to a rainbow diagram of length $k$ (see Figure~\ref{fig1} (b)). 
A straightforward calculation gives 
\begin{equation}
J_0^{(n+k+1)}=
(1+\Delta)^{-k\,\textrm{mod}\,2} (J_0^{(n+1)})^{(-1)^{k}}\Big[\prod\limits_{\alpha=0}^{k-1}
\Big(J_{n+\alpha+1}J_{-(n+\alpha+1)}\Big)^{(-1)^\alpha}\Big]^{(-1)^{k-1}}. 
\end{equation}
Notice that the renormalized coupling $J_0^{\scriptscriptstyle(n+k+1)}$ depends 
on the parity of $k$. The condition for the rainbow diagram to survive one 
SDRG step is still given by Eq.~\eqref{condition-rainbow}, and in the 
logarithmic variables $X_i$ we have
\begin{equation} 
\label{cond-1}
(-1)^{k-1}\Big[X+\sum\limits_{\alpha=0}^{k-1}(-1)^{\alpha-1}
X_{n+\alpha+1}\Big]  \lesssim  \frac{h}{4} - \frac{\ln(1+\Delta)}{2}(k\,\textrm{mod}\,2).  
\end{equation}
The condition~\eqref{cond-1} is the same as for the XX chain apart from the 
parity dependent term $\ln(1+\Delta)/2$. However, this extra term is not  expected to 
change the qualitative behavior of the survival probability. Specifically, in the 
framework of the random walk (compare with Figure~\ref{fig5} (a)), one has that 
the walker is now constrained to stay below $h/4- \log(1+\Delta)/2$ if $k$ is 
odd, and above $-h/4 $ if $k$ is even, i.e., in a strip that  is not symmetric 
around zero (compare with  Figure~\ref{fig5}). It is natural to expect that the decay of 
the survival probability for the walker will remain exponential. 

In stark contrast, the factor $1+\Delta$ in~\eqref{rc-1} has striking consequences 
for the survival probability of the bubble diagrams (see Figure~\ref{fig1} (c)). 
The renormalized coupling $J_0$ due to a bubble diagram of length $2k$ reads 
\begin{equation}
\label{condx}
J_0^{(n+1+2k)}= (1+\Delta)^{-2k}J_0^{(n+1)}  \prod\limits_{\alpha=0}^{2k-1}\Big[
J_{n+\alpha+1}J_{-n-\alpha-1}\Big]^{(-1)^{\alpha+1}}. 
\end{equation}
Using Eq.~\eqref{condx}, the condition Eq.~\eqref{condition-bubble} for 
the survival of the bubble phase can be rewritten as
\begin{equation} 
\label{cond-xx}
X+\sum\limits_{\alpha=0}^{2k-1}(-1)^{\alpha-1}
X_{n+\alpha+1} \gtrsim \frac{h}{4} -k \ln(1+\Delta). 
\end{equation}
In contrast with the non-interacting case, the condition~\eqref{cond-xx} 
depends on the step $k$ of the walker. This means 
that for $\Delta>0$ the survival condition~\eqref{cond-xx} for the walker to be above 
the line $ h/4 - k \ln(1+\Delta)$ is always satisfied for large $k$. 
Physically, this suggests that the bubble phase becomes more and 
more stable as its size increases. However, the short-range singlets 
in the bubble phase do 
not contribute to the entanglement entropy, which explains the 
saturating behavior observed in Figure~\ref{entanglement-int}. On the 
other hand, for $\Delta<0$ one has that for large $k$ the condition~\eqref{cond-xx} is 
never verified. This implies that the bubble phase is suppressed and 
the ground state of the model is in the rainbow phase, with volume-law 
entanglement. 

\section{Conclusions}
\label{sec:conc}

In this paper,  we have  provided evidence of an unusual violation of the area law 
in a random inhomogeneous one-dimensional model. Specifically, we showed 
that in a random inhomogeneous XX chain the ground-state entanglement 
entropy grows with the square root of the subsystem length. We derived 
this result by mapping the SDRG renormalization flow to an alternating 
random walk. The exponent $1/2$ of the entanglement growth can be 
understood from certain survival probabilities of the random walk. 
We also investigated the effect of interactions, considering the random 
inhomogeneous XXZ chain. The unusual area law violation 
is very fragile, and it does not survive when interactions are present. 

It is worth mentioning some research directions for future investigation. 
First, it would be interesting to further study the structure of the renormalization 
group flow in the light of the result~\eqref{theo-res}. For instance, 
it natural to wonder whether ~\eqref{theo-res} might be the staring point for an alternative 
derivation of the Refael and Moore result~\cite{refael-2004} for the entanglement 
entropy in the random XX chain (as well as for the R\'enyi entropies in \cite{fagotti-2011}). 
Another question concerns the fate of  the unusual area-law violation in the limit of weak inhomogeneity. 
In the clean case, i.e., without disorder, using the approach of Ref. \cite{dubail-2016}
it has been shown that the model can be mapped to a CFT in curved spacetime~\cite{laguna-2017,erik}, but 
what happens to this scenario in the presence of randomness is still unknown. 
Also, it would be important to extend the analysis performed in this work to other entanglement-related 
quantities, such as the R\'enyi entropies, the entanglement spectrum and Hamiltonian, 
and the logarithmic negativity. 

Going beyond the behavior of the XX spin-chain, it would be very useful to thoroughly investigate 
the phase diagram of the random inhomogeneous XXZ chain. For instance, 
while for strong inhomogeneity we observed that the entanglement entropy 
has a volume-law scaling for $\Delta<0$, the regime of weak inhomogeneity 
remains unexplored. The most relevant question would be to understand 
whether the volume-law behavior holds true at any value of the inhomogeneity 
or if there is a transition to the expected behavior~\eqref{rsp-ent} taking place 
at finite inhomogeneity. 
Another natural question is whether the mapping between the SDRG flow and the random 
walk allows one to obtain different exotic area-law violations, such as 
a power-law growth of the entanglement with an exponent $\alpha\ne 1/2$. 
A possibility would be to explore the effects of spatially-correlated disorder, 
that in the homogenous case are known to dramatically affect the critical behavior \cite{igloi-rev,igloi-rev-1}.
Moreover,  the nature of the transition between the volume-law and the area-law 
entanglement in the random inhomogeneous XXZ chain  has still to be clarified. 

Finally, an independent, but very timely research direction would be  to understand how the anomalous scaling of the 
ground-state entanglement can affect the out-of-equilibrium behavior of the random inhomogeneous XX chain after 
a (local or global) quantum quench, in particular for the entanglement evolution \cite{cc-05,cc-07l}. 
For instance, for another model with similar anomalous behavior, the spreading of quantum correlations, 
turned out to be very peculiar \cite{anna-2016}.

\subsection*{Acknowledgments} 

SNS has been supported by the Spanish Grant
No. FIS2015-66020-C2-1-P. JRL and GS have been supported by the
Spanish Grants No. FIS2015-69167-C2-1-P, QUITEMAD+ S2013/ICE-2801 and
SEV-2016-0597 of the "Centro de Excelencia Severo Ochoa" Programme.
V.A. acknowledges support from the European Union's Horizon 2020 under
the Marie Sklodowoska-
Curie grant agreement No 702612 OEMBS. 
 P.C. acknowledges support from ERC under Consolidator grant  number 771536 (NEMO). 
Part of this work has been
carried out during the
workshop ``Quantum path''  at  the  Erwin  Schrdinger  International
Institute
for  Mathematics  and  Physics (ESI) in Vienna, and during the
workshop ``Entanglement
in Quantum Systems'' at the Galileo Galilei Institute (GGI) in Florence.


\appendix

\begin{figure}[h!]
\begin{center}
\includegraphics[width=8cm,angle=-90]{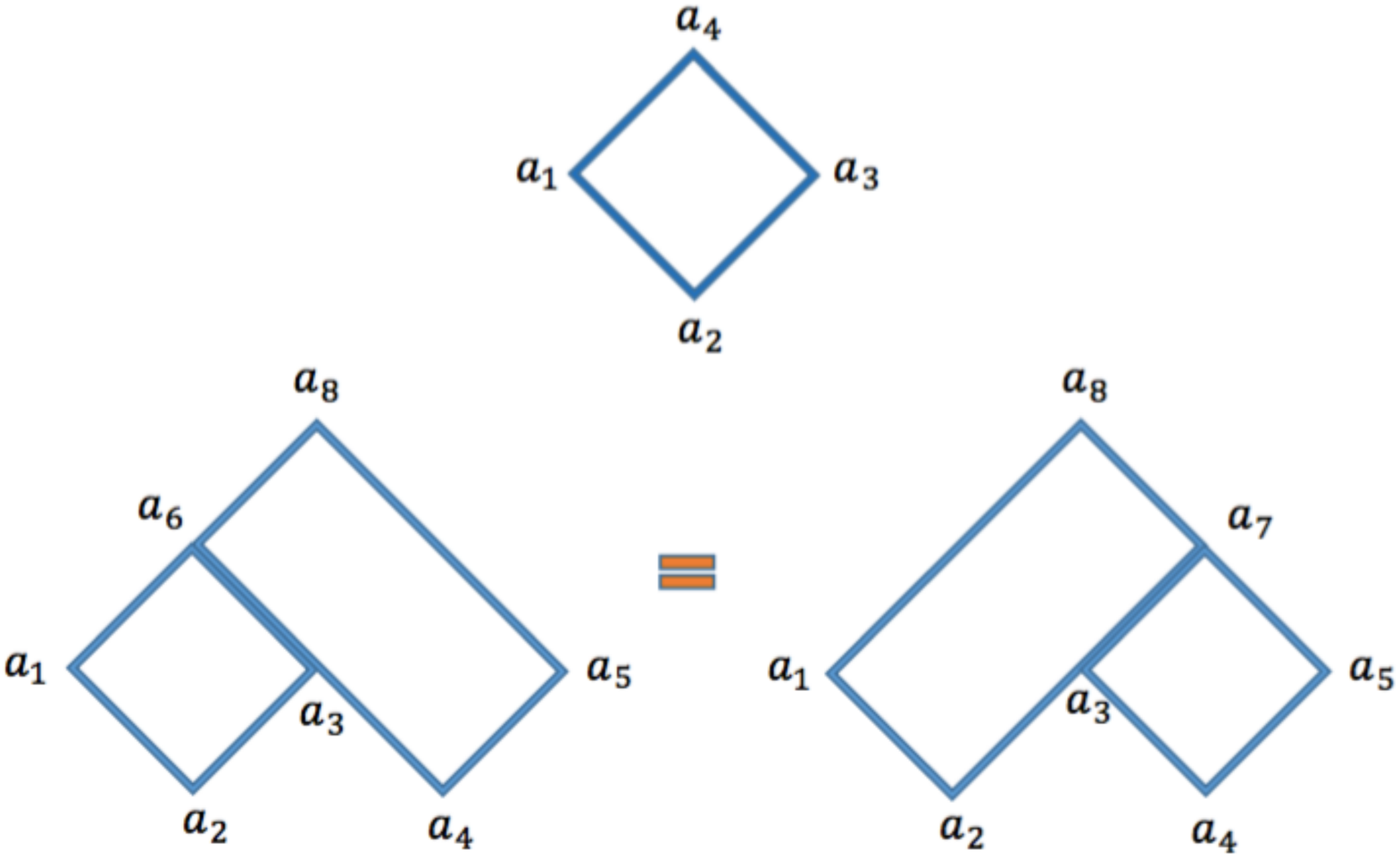}
\end{center}
\caption{Top: geometric representation of the ternary product $a_4 = \{ a_1, a_2, a_3 \}$
(Cf.   Eq.(\ref{rg1})). Fixing the position of the vertices $a_1, a_2$ and $a_3$,  of  the parallelogram determines the
position of the fourth one,  $a_4$.  
Bottom: depiction of the products: $\{ \{ a_1 , a_2 ,   a_3 \} ,   a_4,    a_5 \}  = \{a_6, a_4, a_5 \} = a_8$ and
$ \{ a_1,  a_2, \{  a_3  ,  a_4 ,  a_5 \}  \}  = \{ a_1,  a_2, a_7 \} = a_8$, that is represented by the  
 corner shaped region with the top vertex $a_8$ (Cf. \ref{rg3})). 
}
\label{quasi}
\end{figure} 

\section{SDRG and ternary algebras}
\label{sec-algebra}

In this appendix we propose an  algebraic interpretation of the SDRG procedure
that gives a simple  explanation of the associativity lemma described in section IIB. 
We start by defining an  algebra ${\cal A}$ with a ternary product, that is, 
a map between  three ordered elements into  a fourth one,  
\barray
{\cal A} \times {\cal A} \times {\cal A} & \rightarrow &  {\cal A}
\label{rg1} \\
(a_1, a_2, a_3) &\rightarrow &  a_4 = \{ a_1 , a_2 ,   a_3 \}  \nonumber  \, . 
\earray 
Algebras with a binary product,  say $a_1 \cdot a_2$, are called
associative if the following condition holds:  
$(a_1 \cdot a_2 ) \cdot a_3 = a_1 \cdot (a_2  \cdot a_3)$.  Similarly, 
a ternary algebra is called associative if the triplet product of five elements satisfy \cite{ternary} 
\barray
\{ \{ a_1 , a_2 ,   a_3 \} ,   a_4,    a_5 \} &=  \{  a_1 , \{ a_2,   a_3 ,    a_4 \},     a_5 \}  = \{ a_1,  a_2, \{  a_3  ,  a_4 ,  a_5 \}  \}    \, , 
\label{rg2} 
\earray 
and it is called partially  associative  if the less restrictive condition holds %
\barray
\{ \{ a_1 , a_2 ,   a_3 \} ,   a_4,    a_5 \} & = \{ a_1,  a_2, \{  a_3  ,  a_4 ,  a_5 \}   \}    \, . 
\label{rg3} 
\earray 
An associative algebra is  obviously partially associative
but not vice versa necessarily. 
Graphical representations of the ternary product (\ref{rg1})  and the partial associativity condition (\ref{rg3})  are given in fig.\ref{quasi}. 
%
%
An example of ternary algebra is given by the set on non zero complex numbers, 
${\cal A}= \Cmath - \{ 0 \}$ with  product

\beq
\{ a_1 , a_2 ,   a_3 \}  =  \alpha \frac{ a_1 a_3}{a_2}, \qquad \alpha \neq  0  \, .  
\label{rg4}
\eeq
This definition yields 
%
\barray
\{  a_1 , \{ a_2,   a_3 ,    a_4 \},     a_5 \}& =  &   \frac{a_1 a_3 a_5}{a_2 a_4} ,  \label{rg5} \\ 
\{ \{ a_1 , a_2 ,   a_3 \} ,   a_4,    a_5 \} = \{ a_1,  a_2, \{  a_3  ,  a_4 ,  a_5 \}   \}  & = & 
 \alpha^2 \frac{a_1 a_3 a_5}{a_2 a_4}  , \nonumber 
\earray 
which shows  that ${\cal A}$ is partially associative for any $\alpha \neq 0$, and associative only   for  $\alpha = \pm 1$. 
Notice that the product \eqref{rg4} coincides with the SDRG equation \eqref{rc-1} of the XXZ model, under the identification
$\alpha = 1/( 1+ \Delta)$. Hence the algebra corresponding  to the XX model is associative, a result  that provides  an algebraic derivation of 
the path invariance of the renormalized coupling described in  section IIB.   Indeed, the cartoon 
at the top of Fig.~\ref{fig-theo} (b) 
can be seen to correspond to the products
\beq
\{ J_i  , \{  J_{i+1} ,  J_{i+2} ,  J_{i+3}  \},   J_{i+4} \}  \, , 
\label{rg6}
\eeq 
while  the cartoon at the bottom of Fig.~\ref{fig-theo} (b) correspond to the products 
\beq
\{  \{ J_i ,  J_{i+1},  J_{i+2} \},   J_{i+3}, J_{i+4} \} =  \{  J_i , J_{i+1},  \{ J_{i+2} ,  J_{i+3},  J_{i+4} \} \}  \, . 
\label{rg7}
\eeq 
The equality between  \eqref{rg6} and \eqref{rg7}  amounts to  the associativity of the ternary product. 
Finally, the case $\alpha =-1$, can be shown to correspond
to the SDRG of a free fermion which is equivalent to that of the XX model.


\section{Survival probabilities: Exact results}
\label{sec-calc}

%
\begin{figure}[t]
\includegraphics*[width=1\linewidth]{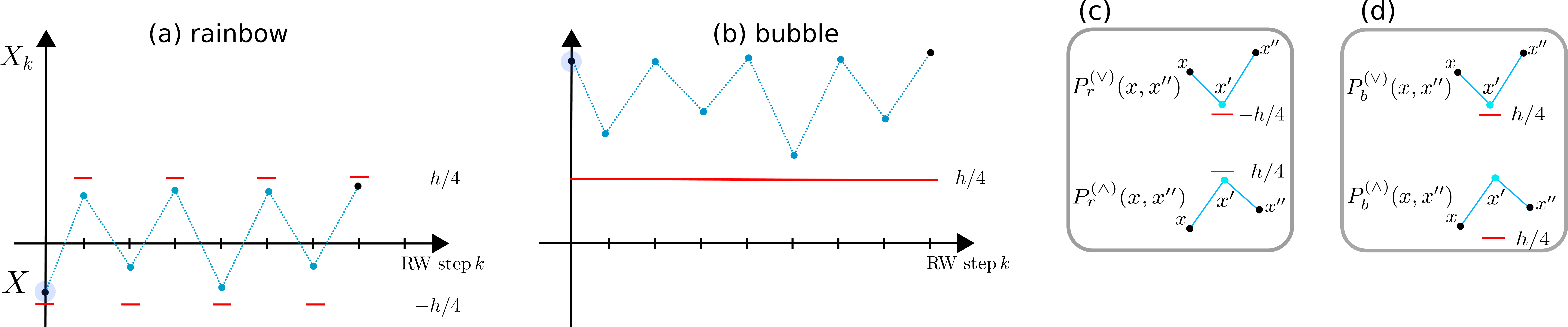}
\caption{ Random walk interpretation of the rainbow diagram (a) and 
 of the bubble diagram (b), same as in Figure~\ref{fig5} (a) and (b). 
 (c) The alternating 
 random walk in (a) is constructed from the two-step survival probabilities 
 $P_r^{\scriptscriptstyle(\vee)}(x,x'')$ or $P_r^{\scriptscriptstyle(\wedge)}
 (x,x'')$. The mid-point $x'$ is integrated over in the interval $[-h/4,\infty]$. 
 $P_r^{\scriptscriptstyle(\vee)}(x,x'')$ corresponds to the probability for the 
 walker to jump from $x$ to $x''$ with the condition $x'>-h/4$. (d) Similar 
 definitions for the two-step probabilities $P_b^{\scriptscriptstyle(\vee)}(x,x'')$ 
 and  $P_b^{\scriptscriptstyle(\wedge)}(x,x'')$ for the random walk in (b). 
}
\label{fig-app}
\end{figure}
%

We now proceed to calculate the survival probability for the alternating 
random walks shown in Figure~\ref{fig-app} (a) and (b) (see also Figure~\ref{fig5}). 
In the following we will employ standard techniques for the random 
walk (see, for instance, Ref.~\onlinecite{majumdar} and 
Ref.~\onlinecite{comtet-2005}). 
The main result of this section will be an exact formula for the 
survival probabilities for the alternating walks in Figure~\ref{fig-app} (a) and (b). 

\subsection{Rainbow diagram}
\label{sec-calc-r}

%
\begin{figure}[t]
\includegraphics*[width=0.8\linewidth]{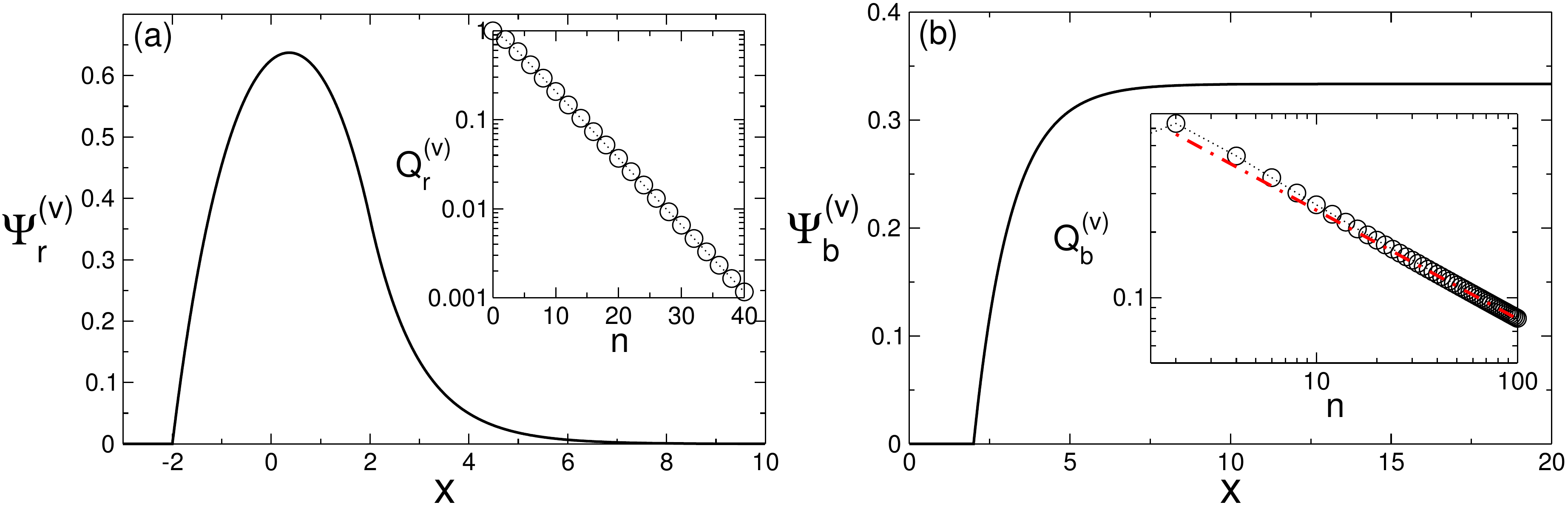}
\caption{ Generating functions $\psi_r^{\scriptscriptstyle(\vee)}(s,x)$ 
 and $\psi_b^{\scriptscriptstyle(\vee)}(s,x)$ for the rainbow  and the 
 bubble survival probabilities. On the $x$-axis $x$ is the initial 
 point of the walker (see Figure~\ref{fig1}). The results are for fixed 
 $s=0.5$ and $h=8$. Notice that $\psi_r^{\scriptscriptstyle(\vee)}$ is 
 zero for $h<-h/4$ and it is vanishing exponentially for $x\gg h/4$.  
 The width of the region where $\psi_r^{\scriptscriptstyle(\vee)}$ is 
 significantly non-zero is $h/2$. Inset: Probabilities $Q^{
 \scriptscriptstyle(\vee)}_r$ that the walker survives {\it at least} 
 for $n$ steps, plotted as a function of $n$. The data are for $h=8$ and 
 $x=0$. Note the logarithmic 
 scale on the $y$-axis signaling the exponential decay with $n$. (b) The 
 same as in (a) for the generating function of the bubble diagrams 
 $\psi_b^{\scriptscriptstyle(\vee)}$. The curve is for $h=8$ and $s=0.25$. 
 Notice that $\psi_b^{
 \scriptscriptstyle(\vee)}$ is exactly zero for $x<h/4$ and it 
 saturates for $x\to\infty$. This reflects that for large $x$ 
 the condition $X_k>h/4$ is relevant only after many steps. Inset: 
 Probabilities $Q^{\scriptscriptstyle(\vee)}_b$ that the walker survives 
 for $n$ steps plotted as a function of $n$. Results are for $h=8$ 
 and $x=3$. Note the logarithmic 
 scale on both axes, signalling a power-law behavior. The dashed-dotted 
 line is the asymptotic result $\propto n^{-1/2}$ for large $n$. 
}
\label{fig7}
\end{figure}
%
We start discussing the alternating random walk in Figure~\ref{fig-app} (a). 
Due to the alternating structure of the random walk, it is first convenient to 
calculate the probability that the walker survives an even number of 
steps. The building blocks are the probabilities $P_r^{\scriptscriptstyle
(\vee)}(x,x'')$ and $P_r^{\scriptscriptstyle(\wedge)}(x,x'')$ (see Figure~\ref{fig-app} (c)
for their pictorial definition). These are the probabilities of surviving two 
steps starting from the initial position  $x$, jumping to a generic intermediate 
point $x'$, and arriving at $x''$. The mid-point $x'$ is integrated over.  
In the definition of $P_r^{\scriptscriptstyle(\vee)}(x,x'')$ 
the intermediate point $x'$ has to satisfy the condition $x'>-h/4$, whereas 
for $P_r^{\scriptscriptstyle(\wedge)}(x,x'')$ one has $x'<h/4$. 
Formally, as it is clear from Figure~\ref{fig-app} (c), the definition 
of $P_r^{\scriptscriptstyle(\vee)}(x,x'')$ is 
\begin{equation}
\label{pv}
P_r^{(\vee)}(x,x'')\equiv \int_{0}^\infty dz\int_0^\infty dz'\delta^{-1} \exp\Big
(-\frac{z+z'}{\delta}\Big)\theta\Big(x-z+\frac{h}{4}\Big)\delta(x''+z-z'-x). 
\end{equation}
In~\eqref{pv} the Heaviside theta function is to ensure the condition $x'>-h/4$.  
After performing the integrals, one obtains 
\begin{equation}
\label{dist}
P^{(\vee)}(x,x'')=\left\{
\begin{array}{cc}
(2 \delta)^{-1}\big(e^{-|x-x''|/\delta}-e^{-(h/2+x+x'')/\delta}\big) & x,x''>-h/4 \\\\
0 &\textrm{otherwise}.
\end{array}\right. 
\end{equation}
To proceed, following the random walk literature~\cite{majumdar}, it is convenient to define 
$G_r^{\scriptscriptstyle(\vee)}(x',x,n)$ as the probability that the walker starts 
from position $x$, it survives $2n$ steps (note the factor $2$), 
and it arrives at $x''$. Due to the markovianity of 
the random walk, $G_r^{\scriptscriptstyle(\vee)}(x',x,n)$ has to 
obey the infinite system of integral equations (one for each value of $n$) 
\begin{equation}
\label{ie1}
G_r^{(\vee)}(x',x,n')=\int_{-\infty}^{\frac{h}{4}}dy G_r^{(\vee)}(x',y,n'-1)
P_r^{(\vee)}(x,y). 
\end{equation}
Eq.~\eqref{ie1} states that the probability for the walker to survive $2n$ steps 
starting from position $x$ and arriving $x'$ is obtained from the product of the 
probability to jump from $x$ to $y$ with that of starting from $y$ and surviving for 
$2n-2$ steps, by summing over all the allowed values of $y$. 

Note that in~\eqref{ie1} 
the integral is in $(-\infty,h/4]$, which prevents us from solving~\eqref{ie1} 
by Fourier transform. To proceed, it is convenient to define the total survival 
probability $Q_r^{\scriptscriptstyle(\vee)}(x,n)$ by integrating over the final 
point of the walker as 
\begin{equation}
\label{qq}
Q^{(\vee)}_r(x,n)\equiv\int_{-\infty}^{\frac{h}{4}}dx' G_r^{(\vee)}(x',x,n). 
\end{equation}
Thus, after using~\eqref{qq}, Eq.~\eqref{ie1} becomes 
\begin{equation}
\label{ie2}
Q_r^{(\vee)}(x,n)=\int_{-\infty}^{\frac{h}{4}}dy Q_r^{(\vee)}(y,n-1)
P_r^{(\vee)}(x,y). 
\end{equation}
To solve the system of equations~\eqref{ie2}, it is convenient to define the 
generating function $\psi^{\scriptscriptstyle(\vee)}_r(x,s)$ as 
\begin{equation}
\label{psi}
\psi^{(\vee)}_r(x,s)\equiv\sum\limits_{m=1}^\infty 
Q_r^{(\vee)}(x,m)s^{m}, 
\end{equation}
where $s$ is a real parameter. The probabilities $Q_r^{\scriptscriptstyle
(\vee)}(x,n)$ are obtained as the coefficients of the Taylor expansion of 
$\psi^{\scriptscriptstyle(\vee)}_r(x,s)$ around $s=0$. After substituting~\eqref{psi} 
in~\eqref{ie2} one obtains the integral equation 
\begin{equation}
\label{ie3}
\psi^{(\vee)}_r(x,s)=s\int_{-\infty}^{\frac{h}{4}}dx'\psi^{(\vee)}_r
(x',s)P_r^{(\vee)}(x,x')+s\int_{-\infty}^{\frac{h}{4}}dx' P_r^{(\vee)}(x,x'). 
\end{equation}
For generic distributions $P_r^{\scriptscriptstyle(\vee)}(x,x')$ it is 
difficult to solve~\eqref{ie3} analytically. However, for the exponential 
distribution in \eqref{dist} the solution of~\eqref{ie3} is straightforward. 
The key ingredient is the identity
\begin{equation}
\label{id}
\big[P_r^{(\vee)}(x,x')\big]''= \frac{1}{\delta^2} \Big[
P_r^{(\vee)}(x,x')-\delta(x-x') \Big]\quad\textrm{with}\,\, 
[f(x)]''\equiv\frac{d^2f}{dx^2}. 
\end{equation}
Thus, after taking the second derivative with respect to 
$x$ in~\eqref{ie3}, and after using~\eqref{id}, one obtains a linear system 
of differential equations as 
\begin{equation}
\label{sys1}
\left\{\begin{array}{ccc}
\delta^2\big[\psi^{\scriptscriptstyle(\vee)}_r\big]''=
(1-s)\psi^{\scriptscriptstyle(\vee)}_r- s  & & -h/4\le x\le h/4\\\\
\delta^2\big[\psi^{\scriptscriptstyle(\vee)}_r(x,s)\big]''=
\psi^{\scriptscriptstyle(\vee)}_r & & x>h/4\\\\
\psi^{\scriptscriptstyle(\vee)}_r=0 & & x<-h/4. 
\end{array}
\right. 
\end{equation}
The solution of~\eqref{sys1} is straightforward. First, one solves 
independently~\eqref{sys1} in the three independent domains 
$(-\infty,-h/4]$, $(-h/4,h/4]$, and $(h/4,\infty)$. After discarding the divergent 
solutions for $x\to\pm\infty$, one has to match the different solutions by 
imposing the continuity of $\psi_r^{\scriptscriptstyle(\vee)}$ at 
$x=\pm h/4$ and of its first derivative $[\psi^{\scriptscriptstyle(\vee)}_r]'$ 
at $x=h/4$. Eventually, one obtains for $-h/4\le x\le h/4$ 
\begin{multline}
\psi^{(\vee)}_r(x,s)=\frac{s}{1-s}
+\frac{s e^{  \sqrt{1-s} h/(4\delta)} }{(-1+e^{\sqrt{1-s}h / 
\delta}+\sqrt{1-s}+e^{\sqrt{1-s} h/ \delta}\sqrt{1-s})(s-1)} 
\\\times\Big[(e^{\sqrt{1-s}h/(2 \delta)}+\sqrt{1-s}-1)e^{\sqrt{1-s}x/
\delta}+(e^{\sqrt{1-s} h/(2 \delta)}+e^{\sqrt{1-s} h /(2 \delta)}
\sqrt{1-s}-1)e^{-\sqrt{1-s}x/ \delta}\Big]. 
\end{multline}
On the other hand, for $x>h/4$, one obtains 
\begin{equation}
\psi^{(\vee)}_r(x,s)=-\frac{s e^{h/(4 \delta)}(-1+e^{
\sqrt{1-s} h/(2 \delta)})^2\sqrt{1-s}e^{-\sqrt{1-s}x / \delta}}
{(-1+e^{\sqrt{1-s}h/\delta}+e^{\sqrt{1-s} h / \delta}\sqrt{1-s})(s-1)}
\end{equation}
Importantly, the generating function, and therefore the probabilities 
$Q_r^{\scriptscriptstyle(\vee)}$, depend only on $h/\delta$ and 
$x/\delta$. This also reflects that the von Neumann entropy is a function 
of the ratio $h/\delta$. 
For completeness we now discuss the result for $\psi^{\scriptscriptstyle
(\wedge)}_r(x,s)$. The probability $P_r^{\scriptscriptstyle(\wedge)}(x,x'')$ 
(see Figure~\ref{fig5} (c)) is now given as 
\begin{equation}
\label{pp}
P_r^{(\wedge)}(x,x'')=\left\{
\begin{array}{cc}
(2 \delta)^{-1}\big(e^{-|x-x''|/ \delta}-e^{(- h/2+x+x'')/
\delta}\big) & x,x''<h/4 \\\\
0 &\textrm{otherwise}.
\end{array}\right. 
\end{equation}
By comparing~\eqref{dist} and~\eqref{pp}, one has that 
$P_r^{\scriptscriptstyle(\wedge)}(x,x'')=P_r^{\scriptscriptstyle(\vee)}(-x,-x'')$. 
One now has to solve the integral equation (the analog of~\eqref{ie3})
\begin{equation}
\label{ie4}
\psi_r^{(\wedge)}(x,s)=s\int_{-\frac{h}{4}}^{\infty}dx'\psi_r^{(\wedge)}
(x',s)P_r^{(\wedge)}(x,x')+s\int_{-\frac{h}{4}}^{\infty}
dx' P_r^{(\wedge)}(x,x'). 
\end{equation}
It is trivial to show that Eq.~\eqref{ie4} is obtained from~\eqref{ie3} 
after the change of variables $x\to-x$. Clearly, this implies that  
\begin{equation}
\psi_r^{(\wedge)}(x,s)=\psi^{(\vee)}_r(-x,s). 
\end{equation}
Finally, we should stress that the $n$-th order coefficient of the Taylor series 
of the generating function $\psi_r^{\scriptscriptstyle(\wedge)}$ around $s=0$ is 
the probabilities $Q_r^{\scriptscriptstyle(\wedge)}$ of the walker to survives for 
at least $2n$ steps within the alternating strip in Figure~\ref{fig5} (a). 
The generating function $\psi_r^{\scriptscriptstyle (\textrm{odd})}$ 
for the probability for the walker to survive an 
odd number of steps $2n+1$ is obtained from $\psi_r^{\scriptscriptstyle(\wedge)}$ 
by performing an extra integration. Formally, one has 
\begin{equation}
\psi_r^{(\textrm{odd})}(x,s)=
\int_{-\frac{h}{4}}^{\frac{h}{4}}dx' \delta^{-1}  e^{-|x-x'| / \delta}
\psi_r^{(\wedge)}(x',s)\theta(x-x'),\quad \textrm{for}\,\,x<\frac{h}{4} 
\end{equation}
whereas, $\psi_r^{\scriptscriptstyle (\textrm{odd})}(x,s)=0$ for $x>h/4$. 
%
%
The survival proability for odd number of steps $Q_r(x,2n+1)$ 
corresponds to the $n$-th coefficient in the series expansion of 
$\psi_r^{\scriptscriptstyle(\textrm{odd})}(x,s)$ around $s=0$. 

It is interesting to investigate the behavior of $\psi_r^{\scriptscriptstyle
(\vee)}$ as a function of the initial point of the walker $x$. This is 
shown in Figure~\ref{fig7} (a) for fixed $s=0.5$, $h=8$, and $\delta=1$. 
$\psi_r^{\scriptscriptstyle(\vee)}$ is zero for $x<-h/4$ and it 
is vanishing exponentially for $x\to\infty$. Similar behavior is found 
for different values of $s$. The inset in the Figure shows the 
probability $Q_r^{\scriptscriptstyle(\vee)}$ for the walker to 
survive $n$ steps, as a function of $n$. Clearly, it decays 
exponentially (note the logarithmic scale on 
the $y$-axis). 

\subsection{Bubble diagram}
\label{sec-b}

We now derive analytically the probability for the walker to remain above the 
line $h/4$ (see Figure~\ref{fig-app} (b)). As in the previous section, 
we can define the generating function $\psi_b^{\scriptscriptstyle(\vee)}
(x,s)$ for the probability to survive an even number of steps $n$. 
The building block two-steps probability $P_b^{\scriptscriptstyle(\vee)}(x,x'')$ 
is defined in Figure~\ref{fig-app} (d), and it is given as 
\begin{equation}
\label{dist2}
P_s^{(\vee)}(x,x'')=\left\{
\begin{array}{cc}
(2\delta)^{-1}\big(e^{-|x-x''|/ \delta}-e^{(h/2-x-x'')/ \delta}\big) & x,x''>h/4 \\\\
0 &\textrm{otherwise}.
\end{array}\right. 
\end{equation}
Notice that $P_b^{\scriptscriptstyle(\vee)}$ is obtained from 
$P_r^{\scriptscriptstyle(\vee)}$ after sending $h\to-h$, as it clear from 
comparing Figure~\ref{fig-app} (c) and (d). The equation for 
$\psi_b^{\scriptscriptstyle(\vee)}(x,s)$ reads 
\begin{equation}
\label{b-gen}
\psi_b^{(\vee)}(x,s)=s\int_{\frac{h}{4}}^\infty dx'
P_b^{(\vee)}(x,x')\psi_s^{(\vee)}(x',s)+
s\int_{\frac{h}{4}}^\infty dx'P_b^{(\vee)}(x,x'). 
\end{equation}
As in the previous section, Eq.~\eqref{b-gen} can be reduced to a system of differential 
equations. Specifically, $\psi_b^{\scriptscriptstyle(\vee)}$ is obtained 
by solving the system   
\begin{equation}
\label{sys3}
\left\{\begin{array}{ccc}
\psi_b^{\scriptscriptstyle(\vee)}(x,s)=0 & & x \leq h/4\\\\ 
\delta^2[\psi^{\scriptscriptstyle(\vee)}_b]''=(1-s)
\psi_b^{\scriptscriptstyle(\vee)}- s & & x> h/4.\\\\
\end{array}
\right. 
\end{equation}
In contrast with Eq.~\eqref{sys1}, 
the domain of the system~\eqref{sys3} is composed of the two half-infinite 
intervals $(-\infty,h/4]$ and $[h/4,\infty)$. The solution for $x>h/4$ is given as 
\begin{equation}
\psi_b^{(\vee)}=-\frac{s}{1-s}e^{-\sqrt{1-s}(x-h/4)/\delta}+\frac{s}{1-s}
\end{equation}
%
%
%
%
%
%
The strategy to derive the generating function 
$\psi_b^{\scriptscriptstyle(\wedge)}$ is similar, and we do not report the 
calculation, quoting only the final result. One obtains 
\begin{align}
\label{wedge1}
& \psi_b^{(\wedge)}=\frac{s}{1-s}+\frac{se^{\sqrt{1-s} (h/4-x) / \delta}}
{(1+\sqrt{1-s})(s-1)},&\textrm{for}\,\,x>h/4\\
\label{wedge2}
& \psi_b^{(\wedge)}=\frac{se^{(-h/4+x)/ \delta}}
{(1+\sqrt{1-s})\sqrt{1-s}},&\textrm{for}\,\,x\le h/4. 
\end{align}
In the limit $x\to\infty$ one has $\psi_b^{\scriptscriptstyle
(\wedge)}=s/(1-s)$. This reflects that if the starting point of the 
walker is at $x\to\infty$, it remains above the line $h/4$ 
for an infinite number of steps. 

The probability $Q_b^{\scriptscriptstyle(\wedge)}$ that the walker survives for at least 
$2n$ steps is the $n$-th coefficient of the Taylor expansion of 
$\psi_b^{\scriptscriptstyle(\wedge)}$ around $s=0$.  Notice that 
for the bubble diagram, the probability that the walker survives 
at least for $2n+1$ steps is the same as that for surviving $2n$ 
steps. This is clear from Figure~\ref{fig7} (b), and it is due to 
the fact that the random walk is alternating. 
%

We now discuss the structure of $\psi_b^{\scriptscriptstyle
(\vee)}$. A similar behavior is observed for $\psi_b^{\scriptscriptstyle(\wedge)}$. 
The result is shown in Figure~\ref{fig7} (b). The data are for $h=8$ and $s=0.25$. 
Clearly, the generating function vanishes for $x<h/4$, while it saturates to $s/(1-s)$ 
in the limit $x\to\infty$. The inset in the Figure shows the survival probabilities 
$Q_b^{\scriptscriptstyle(\vee)}(x,n)$ for the walker to start from $x$ and survive for 
at least $n$ steps. In contrast with the survival probability for the rainbow, which 
decays exponentially (see Figure~\ref{fig7} (a)), now the decay for large $n$ 
is power law as 
\begin{equation}
\label{prev}
Q_b^{(\vee)}(x,n)\propto n^{-\frac{1}{2}}. 
\end{equation}
Using Eq.~\eqref{prev} it is straightforward to derive the probability 
$P_b^{\scriptscriptstyle(\vee)}$ that the walker survives {\it exactly} 
$n$ steps as 
\begin{equation}
P_b^{(\vee)}\equiv Q_b^{(\vee)}(n)-Q_b^{(\vee)}(n+1). 
\end{equation}
Since $Q_b^{\scriptscriptstyle(\vee)}(n+1)=Q_b^{\scriptscriptstyle(\vee)}(n)$ 
for any $n$ odd, one has that $P_b^{\scriptscriptstyle(\vee)}(n)=0$ for any 
$n$ odd. Oppositely, one has that $P_b^{\scriptscriptstyle(\wedge)}(n)=0$ for 
$n$ even. Finally, using~\eqref{prev}, one has that $P_b^{\scriptscriptstyle(\vee)}$ 
(and $P_b^{\scriptscriptstyle(\wedge)}$) for large $n$ decays as 
\begin{equation}
P^{(\vee)}_b(n)\equiv Q^{(\vee)}_b(x, n)-Q^{(\vee)}_b(x,n+1)\propto n^{-\frac{3}{2}}. 
\end{equation}
%

\subsection{Asymptotic behavior of the length of the bubble diagrams} 
\label{sec-asy}

%
\begin{figure}[t]
\includegraphics*[width=0.55\linewidth]{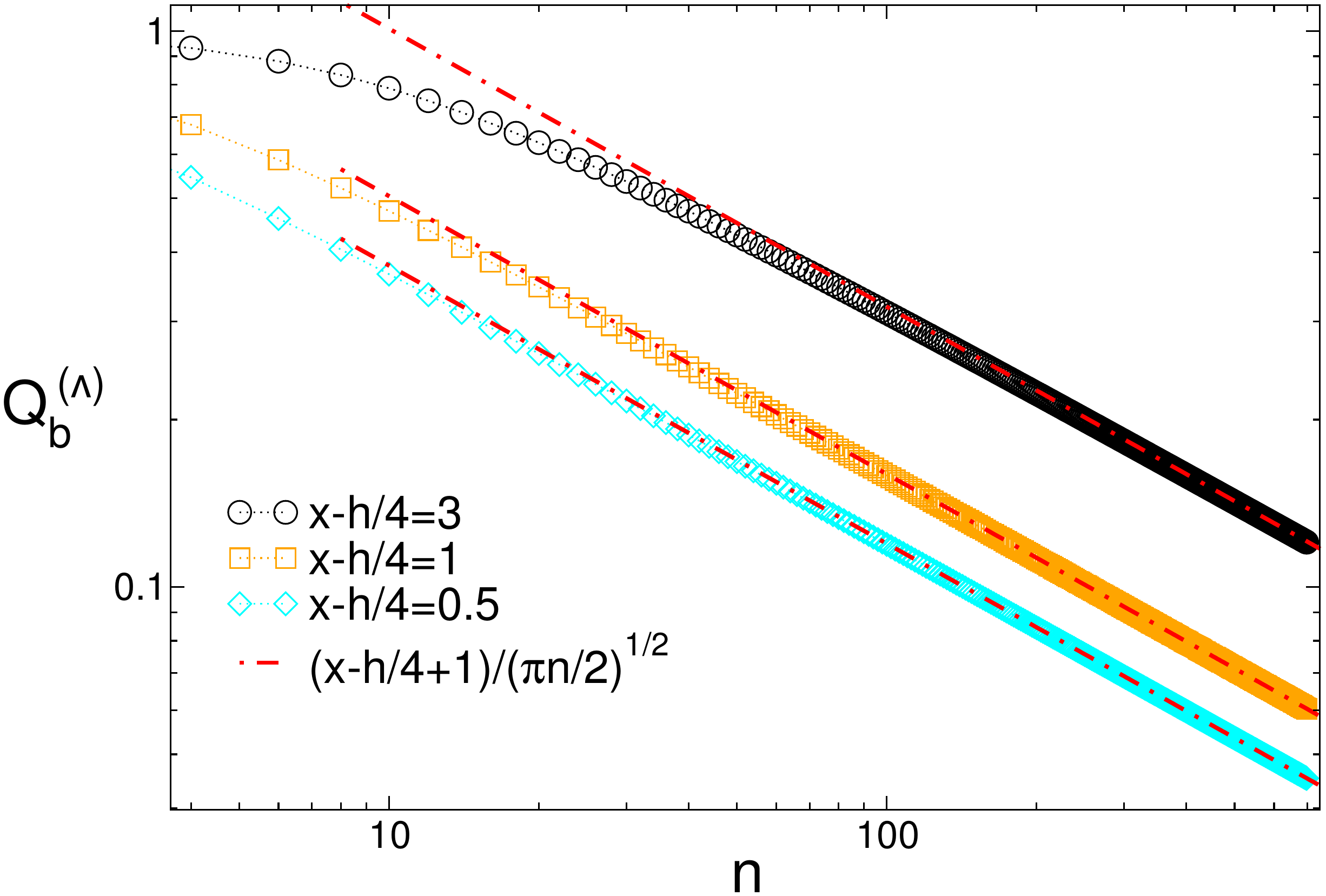}
\caption{Survival probabilities $Q_b^{\scriptscriptstyle(\wedge)}$ 
 for the bubble diagrams. Here $Q_b^{\scriptscriptstyle(\wedge)}(n)$ 
 is the probability to have a bubble diagram of at least $n$ sites. 
 In the random walk language (see Figure~\ref{fig5} (b)) this corresponds 
 to the probability of the walker to start from the initial point $x$ and 
 to stay above the line $h/4$ for at least $n$ steps. The probabilities 
 depend only on the combination $x-h/4$ (different symbols in the Figure). 
 Here we fixed $\delta=1$. 
 The dashed-dotted lines are the asymptotic results in the limit $n\to\infty$. 
 Notice that one has $Q^{\scriptscriptstyle (\wedge)}_b=(x-h/4+1)/(\pi n/2)^{1/2}$ 
 for $n\to\infty$.
}
\label{fig7-a}
\end{figure}
%

The asymptotic behavior as $n^{-1/2}$ of $Q^{\scriptscriptstyle
(\wedge)}_b(x,n)$ (and of $Q_b^{\scriptscriptstyle(\wedge)}$) can be 
obtained analytically by studying the coefficients of the Taylor series 
of $\psi^{\scriptscriptstyle
(\wedge)}_s(x,s)$ around $s=0$. Here we focus on the generating function  
for $x>h/4$. First, it is convenient to rewrite the generating 
function~\eqref{wedge1} as 
\begin{equation}
\label{f-asy}
\psi_b^{(\wedge)}(x,s)=\frac{s}{1-s}-se^{\sqrt{1-s}(h/4-x)/\delta}
\Big[\frac{1}{1+(1-s)^\frac{1}{2}}+\frac{1}{1-s}
-\frac{1}{(1-s)^\frac{1}{2}}\Big]. 
\end{equation}
One can check that the first term in the square brackets in~\eqref{f-asy} 
gives a subleading contribution in the limit $n\to\infty$, and it can be 
neglected. Now we focus on the second term in the brackets. 
It is useful to observe that 
\begin{equation}
\label{asy2}
\frac{e^{-a\sqrt{1-s}}}{1-s}=-a\sum\limits_{k=0}^\infty \frac{a^{2k}(1-s)^{k-1/2}}{(2k+1)!}+
\sum\limits_{k=0}^\infty \frac{a^{2k}(1-s)^{k-1}}{(2k)!}, \quad\textrm{with}\,\,
a\equiv \frac{1}{\delta}\big(x-\frac{h}{4}\big).
\end{equation}
The first term in~\eqref{asy2} can be rewritten as 
\begin{equation}
\label{asy3}
-a\sum\limits_{k=0}^\infty \frac{a^{2k}(1-s)^{k-1/2}}{(2k+1)!}
\equiv\sum_{n=0} c_n s^n, \quad\textrm{with}\, 
c_n\equiv -\sum\limits_{k=0}^\infty
\frac{a^{2k+1}\Gamma(n-k+1/2)}{\Gamma(-k+1/2)\Gamma(n+1)\Gamma(2k+2)}. 
\end{equation}
%
%
%
Here we are interested in the limit $n\to\infty$ of $c_n$. In this limit 
one can expand the terms $\Gamma(n-k+1/2)$ and $\Gamma(n+1)$ in~\eqref{asy3} 
to obtain  
\begin{equation}
\label{f-asy-1}
c_n\to  
-\sum_{k=0}^\infty\frac{a^{2k+1}n^{-k-1/2}}{\Gamma(-k+1/2)\Gamma(2k+2)}=
-\textrm{Erf}\Big(\frac{a}{2\sqrt{n}}
\Big)\to-\frac{a}{\sqrt{\pi n}}.
\end{equation}
%
%
%
In the last step in~\eqref{f-asy-1} we used the expansion of the 
error function $\textrm{Erf}(x)$ around $x=0$. On the other hand, 
the second sum in~\eqref{asy2} gives 
\begin{equation}
\sum\limits_{k=0}^\infty \frac{a^{2k}(1-s)^{k-1}}{(2k)!}
\equiv\sum_{n=0}^\infty c'_n s^n,\quad\textrm{with}\,
c'_n\equiv\sum
\limits_{k=0}^\infty\frac{a^{2k}\Gamma(n-k+1)}{\Gamma(-k+1)\Gamma(n+1)\Gamma(2k+1)}. 
\end{equation}
The factor $\Gamma(-k+1)$ implies that only the term with $k=0$ 
is nonzero. The result cancels out with the term $s/(1-s)$ in~\eqref{f-asy}. 

A similar analysis can be performed for the last term in~\eqref{f-asy}. 
One obtains  
\begin{equation}
\label{last}
\frac{se^{-a\sqrt{1-s}}}
{(1-s)^\frac{1}{2}}\equiv\sum\limits_{n=0}^\infty c_n'' s^n,\quad
\textrm{with}\, c''_n\to\frac{1}{\sqrt{\pi n}}. 
\end{equation}
Notice that the right hand side in~\eqref{last} does not depend on $a$. 
The asymptotic behavior of $Q_b^{\scriptscriptstyle(\vee)}$ for large $n$ 
is obtained by putting together~\eqref{last} and~\eqref{f-asy-1}. A 
similar approach can be used to derive the asymptotic behavior of 
$Q_b^{\scriptscriptstyle(\vee)}(x,n)$. Our final formulas read 
\begin{equation}
\label{asy-res}
Q_b^{(\wedge)}(x,n)\propto\Big(x-\frac{h}{4}+1\Big)\frac{\sqrt{2}}{\delta\sqrt{\pi n}}, 
\quad\quad Q_b^{(\vee)}(x,n)\propto\Big(x-\frac{h}{4}\Big)\frac{\sqrt{2}}{\delta\sqrt{\pi n}}. 
\end{equation}
The factor $\sqrt{2}$ in~\eqref{asy-res} takes into account that the 
$n$-th order coefficient in the expansion of the generating functions 
around $s=0$ is the probability of surviving at least $2n$ steps. 

Finally, we provide some numerical checks of the validity of~\eqref{asy-res}. 
The results are reported in Figure~\ref{fig7-a} for 
$Q^{\scriptscriptstyle(\wedge)}_b(x,n)$. The Figure shows $Q^{\scriptscriptstyle
(\wedge)}_b(x,n)$ versus $n$. The different 
symbols correspond to different values of $x-h/4$. The dashed-dotted lines 
are the asymptotic results obtained from~\eqref{asy-res}, 
and they perfectly describe the numerical data. 

%
\begin{figure}[t]
\includegraphics*[width=0.8\linewidth]{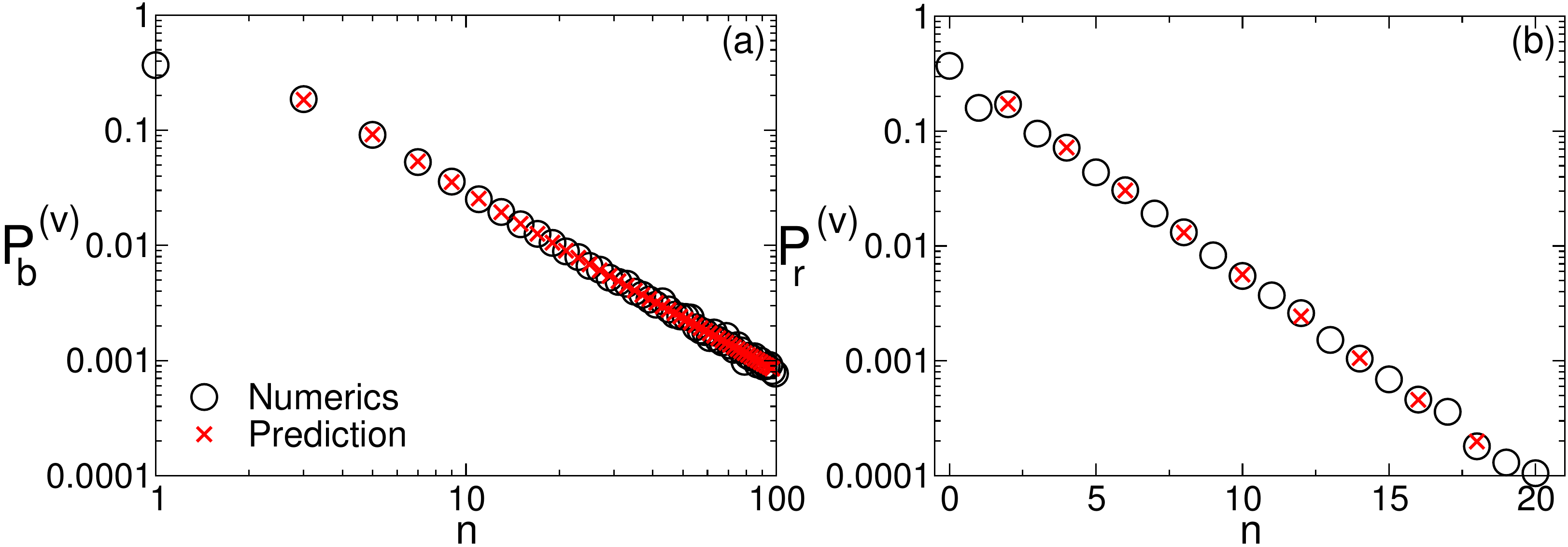}
\caption{The alternating random walk: Numerical checks. The two 
 panels show the walker survival probabilities $P^{\scriptscriptstyle(\vee)}_b(n)$ and 
 $P^{\scriptscriptstyle(\vee)}_r(n)$ (panels (a) and (b), respectively). 
 Here $P^{\scriptscriptstyle(\vee)}_b(n)$ 
 is defined as the probability that the walker starts from the initial 
 point $x$ and remains above the line $h/4$ (see Figure~\ref{fig5}) 
 {\it exactly} for $n$ steps. Notice that $P_b^{\scriptscriptstyle(\vee)}(2k)=0$ 
 for any $k$ due to the alternation. $P^{\scriptscriptstyle(\vee)}_r(n)$ is the probability 
 of the walker to start from $x$ and stay confined in the staggered 
 strip $[-h/4,h/4]$ (see Figure~\ref{fig5} (b)) for $n$ steps. 
 The circles in the panels are obtained by numerically simulating the 
 alternating random walk and correspond to an average over $\sim1000$ 
 realization of the walk. We used $x=2$ in (a) and $x=0$ in (b) 
 and $h=4$. In both panels the crosses are analytical results. In (b) 
 only results for even $n$ are reported. 
}
\label{rw_check}
\end{figure}
%

\section{Random walk survival probabilities: Numerical simulations}
\label{sec-bench}

In this section we provide some numerical checks of the analytical results for the 
walker survival probabilities $P^{\scriptscriptstyle(\vee)}_r(n)$ and 
$P^{\scriptscriptstyle(\vee)}_b(n)$. The former is the probability that 
the alternating random walk is confined in the strip 
$[-h/4,h/4]$ (see Figure~\ref{fig-app} (a)) for exactly $n$ steps, 
whereas the latter is the probability that the walk remains above the 
line $h/4$ for exactly $n$ steps (see Figure~\ref{fig-app} (b)). 
The analytical results for the probabilities are obtained as 
\begin{align}
\label{pr}
& P_b^{(\vee)}(n)=Q_b^{(\vee)}(n)-Q^{(\vee)}_b(n+1),\\
\label{pr1}
& P_r^{(\vee)}(n)=Q_r^{(\vee)}(n)-Q^{(\vee)}_r(n+1), 
\end{align}
where $Q_b^{\scriptscriptstyle(\vee)}$ and $Q_r^{\scriptscriptstyle(\vee)}$ 
are obtained from the generating functions $\psi_b^{\scriptscriptstyle(\vee)}$ 
and $\psi_r^{\scriptscriptstyle(\vee)}$. 
To benchmark Eq.~\eqref{pr} here we present numerical data obtained by 
simulating directly the alternating random walk (cf.~\eqref{theo-res}) 
that describes the SDRG flow of the couplings. 
Our results are shown in Figure~\ref{rw_check}. Panel (a) shows 
$P^{\scriptscriptstyle(\vee)}_b(n)$ plotted as function 
of $n$. The data are for 
fixed $h=4$ and initial point of the walker $x=h/4$. The data are 
averaged over $\sim1000$ realizations of the random walk. Notice that,  
due to the alternating structure (see Figure~\ref{fig-app}), 
one has that $P^{\scriptscriptstyle(\vee)}_b(2n)=0$, as it is clear from 
Figure~\ref{fig-app} (b). The crosses Figure~\ref{rw_check} are 
the theoretical predictions using~\eqref{pr1}. Panel (b) in the Figure 
focuses on the random walk in Figure~\ref{fig-app} (a). For simplicity we focus on the 
case with $h=4$ and $x=0$. The survival probabailities decay exponentially 
with $n$ (note the logarithmic scale on the $y$-axis). The crosses are 
now the theory predictions calculated from~\eqref{pr}. We only 
show results for even $n$, although $P_r(n)$ for $n$ odd is nonzero. 
Clearly, the analytical results are in perfect agreement with the 
numerical data.







\end{document}